%% file: acl_latex.tex
\documentclass[11pt]{article}

\usepackage[final]{acl}

\usepackage{times}
\usepackage{latexsym}
\usepackage{multirow}
\usepackage[T1]{fontenc}

\usepackage[utf8]{inputenc}

\usepackage{microtype}

\usepackage{inconsolata}

\usepackage{subcaption}
\usepackage{tikz}
\usepackage{fontawesome5}
\usepackage{booktabs}
\usepackage{graphicx}
\usepackage{pifont}
\usepackage{makecell} 
\usepackage{xcolor}
\usepackage{colortbl}
\usepackage{amsmath}
\usepackage{amssymb}
\usepackage{amsthm}
\usepackage{bm}
\usepackage[most]{tcolorbox}
\usepackage{enumitem}
\usepackage{placeins}
\usepackage{pgfplots} 
\pgfplotsset{compat=1.18}

\theoremstyle{plain}

\usepackage{titlesec}    
\usepackage{titletoc}    

\title{\raisebox{-0.7em}{\includegraphics[height=2em]{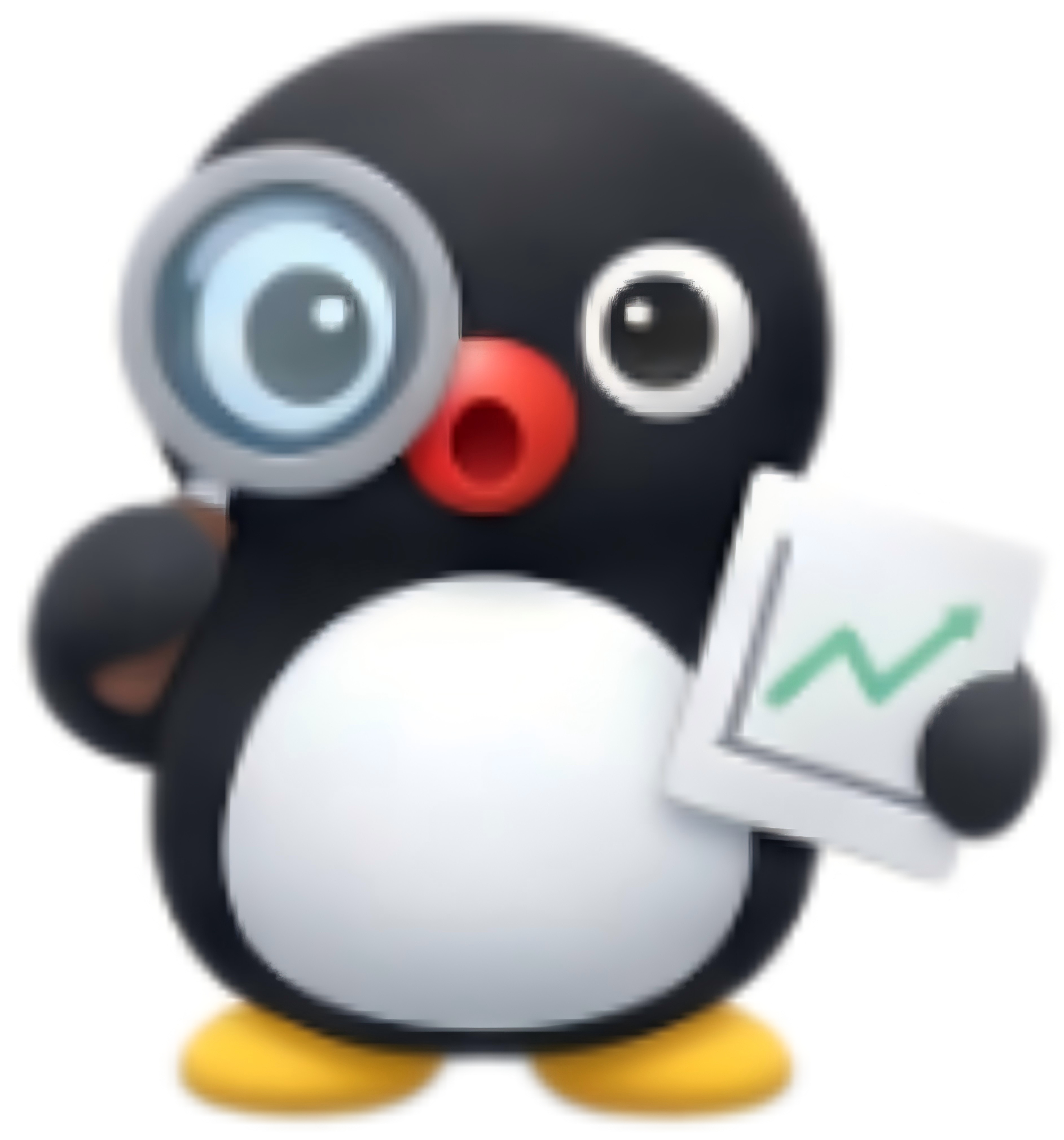}} QuantaAlpha: An Evolutionary Framework for 
\\LLM-Driven Alpha Mining}

\vspace{5mm}

\author{
\textbf{Jun Han\textsuperscript{1*}},
 \textbf{Shuo Zhang\textsuperscript{2*}},
 \textbf{Wei Li\textsuperscript{1*}},
 \textbf{Yifan Dong\textsuperscript{1}},
 \textbf{Tu Hu\textsuperscript{2}},
 \textbf{Yumo Zhu\textsuperscript{1}},
\\
 \textbf{Xiaomin Yu\textsuperscript{2}},
 \textbf{Xin Guo\textsuperscript{1}},
 \textbf{Zhaowei Liu\textsuperscript{1}},
 \textbf{Kunyi Wang\textsuperscript{2}},
  \textbf{Jingping Liu\textsuperscript{3}},
  \textbf{Tianyi Jiang\textsuperscript{4}},
\\
 \textbf{Ruichuan An\textsuperscript{4}},
 \textbf{Sen Hu\textsuperscript{2,4}},
 \textbf{Zhi Yang\textsuperscript{1\dag}},
 \textbf{Ronghao Chen\textsuperscript{2,4\dag}},
 \textbf{Huacan Wang\textsuperscript{2\dag}}\vspace{2mm}
\\
 \textsuperscript{1}SUFE,\,
 \textsuperscript{2}QuantaAlpha,\,
 \textsuperscript{3}SYSU,\,
 \textsuperscript{4}PKU \vspace{2mm}
 \\
\small {
    \textbf{{*}These authors contributed equally to this work.}
}
\\
 \small{
   \textbf{\dag Correspondence:} 
   \href{mailto:yangzhi_sufe@outlook.com}{yangzhi\_sufe@outlook.com},
\href{mailto:chenronghao@alumni.pku.edu.cn}{chenronghao@alumni.pku.edu.cn},
\href{mailto:wanghuacan17@mails.ucas.ac.cn}{wanghuacan17@mails.ucas.ac.cn}
 }
}
\newcommand{\method}{QuantaAlpha}

\begin{document}
\maketitle
\begin{center}
    \large
    \faGithub\hspace{6pt}\href{https://github.com/QuantaAlpha/QuantaAlpha}{\texttt{\color{black}https://github.com/QuantaAlpha/QuantaAlpha}}
\end{center}

\begin{abstract}
Financial markets are noisy and non-stationary, making alpha mining highly sensitive to backtest noise and regime shifts. While recent agentic frameworks improve automation, they often lack controllable multi-round search and reliable reuse of validated experience. To address these challenges, we propose \textbf{QuantaAlpha}, an evolutionary alpha mining framework that treats each end-to-end mining run as a trajectory and improves factors via trajectory-level mutation and crossover. QuantaAlpha localizes suboptimal steps for targeted revision and recombines complementary high-reward segments to reuse effective patterns, enabling structured exploration and refinement across iterations. During factor generation, it enforces semantic consistency across hypothesis, factor expression, and executable code, and constrains the complexity and redundancy of the generated factor to mitigate crowding. Extensive experiments on CSI 300 show consistent gains over strong baselines and prior agentic systems. Using GPT-5.2, QuantaAlpha achieves an IC of 0.0472 with ARR of 4.68\% and MDD of 11.8\%. Moreover, factors mined on CSI 300 transfer effectively to CSI 500 and the S\&P 500, delivering about 40.28\% and 19.1\% cumulative excess return over four years, respectively, which indicates strong robustness under market distribution shifts.

\end{abstract}
\begin{figure*}[htbp]
\centering
\begin{subfigure}[t]{0.55\textwidth}
    \centering
    \includegraphics[height=0.28\textheight,keepaspectratio]{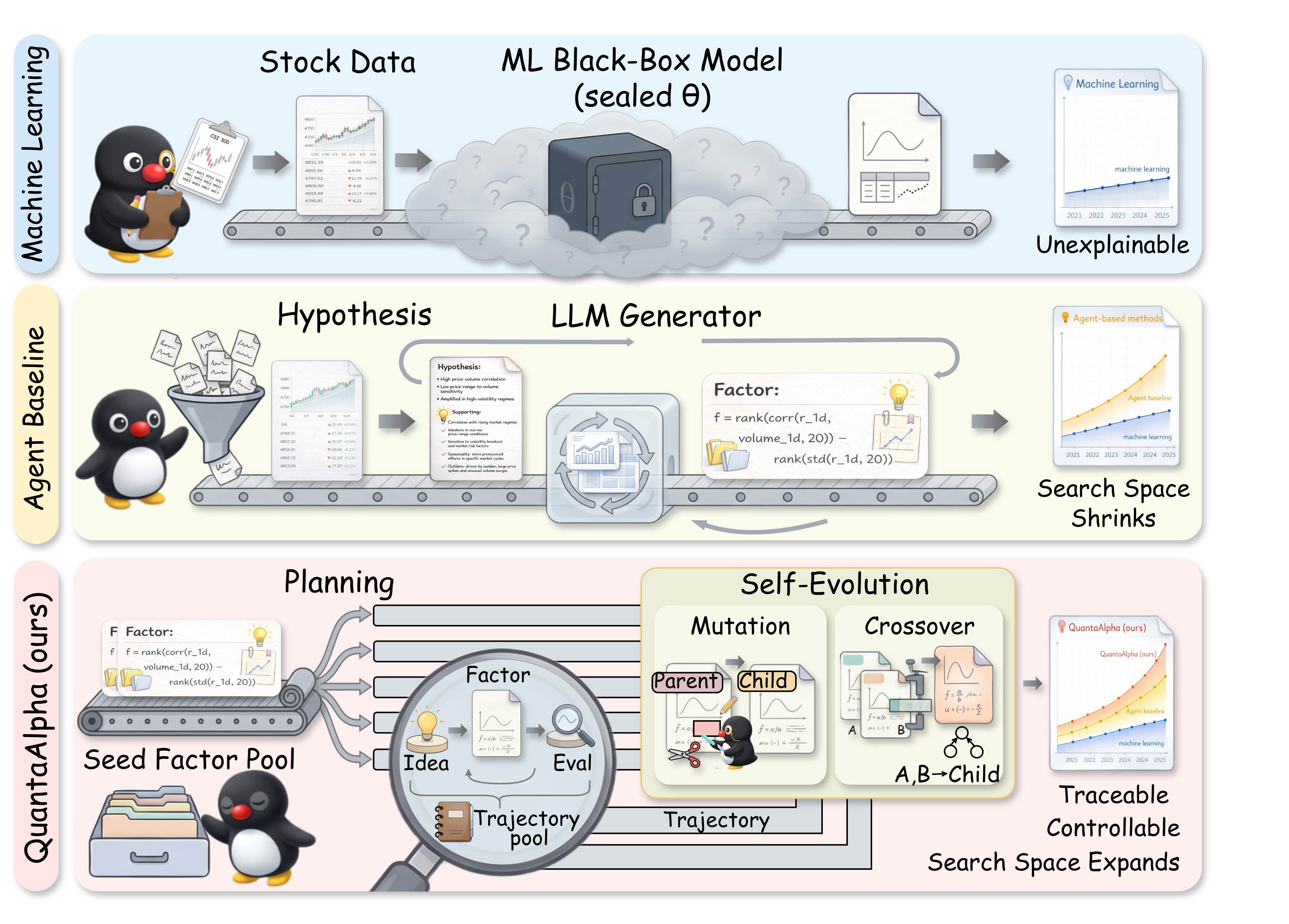}
    \phantomcaption
    \label{fig:intro_overview}
\end{subfigure}
\hspace{0.02\textwidth}
\begin{subfigure}[t]{0.41\textwidth}
    \centering
    \includegraphics[height=0.28\textheight,keepaspectratio]{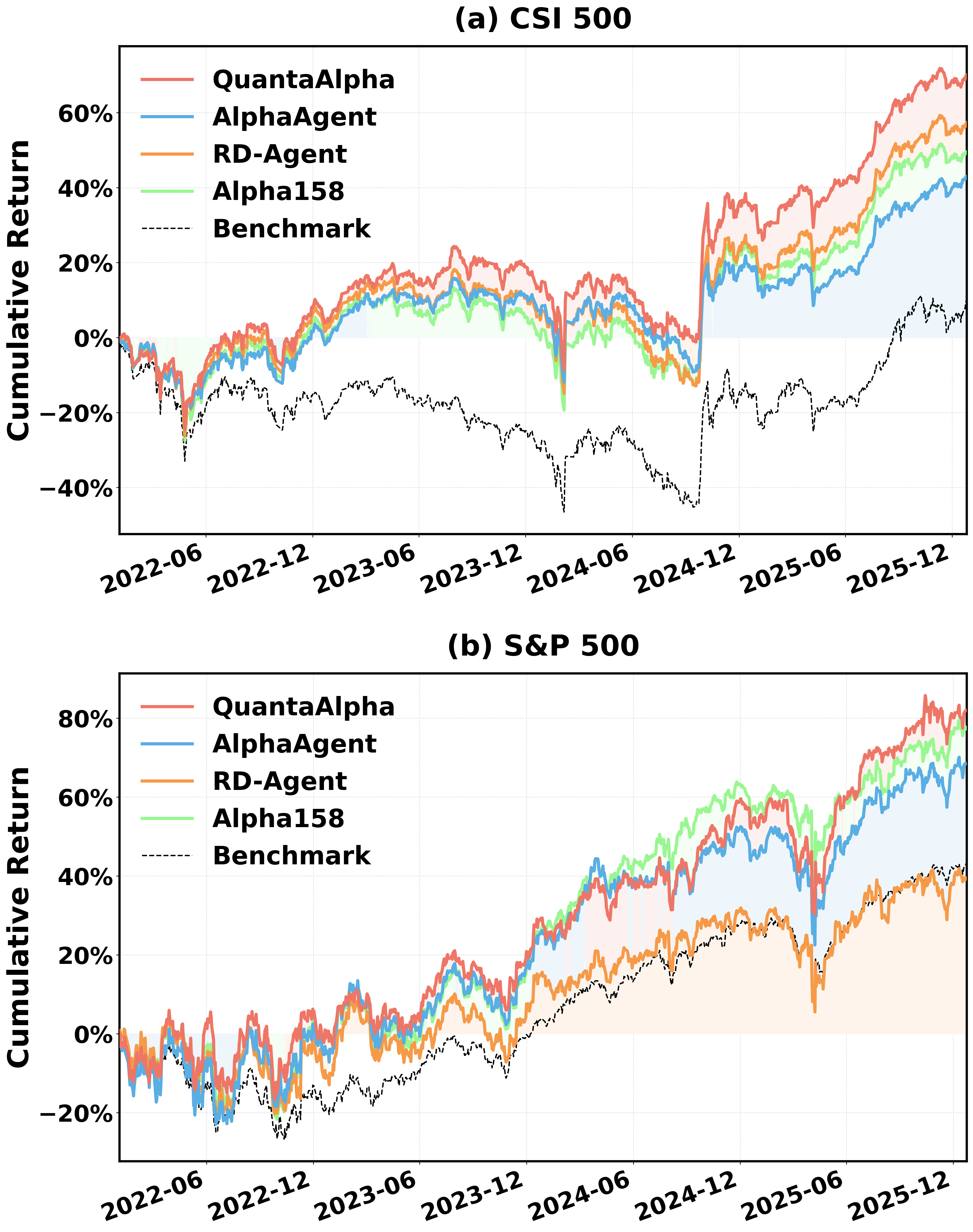}
    \phantomcaption
    \label{fig:cumulative_returns}
\end{subfigure}

\caption{\footnotesize Overview and empirical performance of QuantaAlpha. 
Left: QuantaAlpha improves alpha discovery through trajectory-level self-evolution. 
Right: Cumulative excess returns of different approaches on CSI 500 and S\&P 500.}
\label{fig:intro_combined}
\end{figure*}
\newpage


\input{main_section/1.introduction}

\input{main_section/2.related_work}

\input{main_section/3.problem_formulation}

\input{main_section/4.method}
\input{main_section/5.experiments}

\input{main_section/6.conclusion}

\input{main_section/7.acknowledgement}


\bibliography{custom}

\newpage
\appendix
\onecolumn

\centerline{\maketitle{\textbf{SUMMARY OF THE APPENDIX}}}

This appendix contains additional details for the  \textbf{\textit{``QuantaAlpha: An Evolutionary Framework for LLM-Driven Alpha Mining''}}. The appendix is organized as follows:

\startcontents[appendices]
\section*{Appendix Table of Contents}
\printcontents[appendices]{}{0}{\large}

\input{appendix/1.app_1}
\input{appendix/2.app_2}
\input{appendix/3.app_3}

\input{appendix/4.app_4}

\end{document}

%% file: main_section/1.introduction.tex
\section{Introduction}
Financial markets are high-dimensional, non-stationary stochastic systems, where returns exhibit heavy tails \cite{fama1965behavior}, time-varying volatility \cite{engle1982autoregressive}, and cross-sectional dependence \cite{pesaran2021general}. Quantitative investment therefore relies on alpha mining to extract predictive signals from noisy data. Recently, large language models (LLMs) \cite{lopez2023can} and LLM-based agent frameworks \cite{xiao2024tradingagents,zhang2024multimodal} have been introduced into factor research. By leveraging reasoning and code generation capabilities, these systems can automate factor construction and iteratively refine candidates through backtesting feedback.

Representative agent frameworks for alpha mining mimic human quantitative researchers by iteratively performing hypothesis generation, factor construction, and backtesting-based refinement. Under this paradigm, AlphaAgent \cite{tang2025alphaagent} imposes explicit regularization to mitigate crowding and alpha decay, while RD-Agent \cite{li2025r} enables full-stack automation through coordinated research and development agents for joint factor–model optimization. These frameworks reduce trial-and-error costs while preserving interpretability, making large-scale exploration practical.



However, under real-market non-stationarity and low signal-to-noise ratios, existing systems still face three limitations. \textit{(i) Fragile controllability:} Iterative refinement is driven by noisy backtest feedback, which can induce semantic drift and steer updates toward spurious correlations, gradually deviating from the intended economic mechanism. \textit{(ii) Limited trustworthiness:} Many methods rely on stochastic re-generation conditioned on transient contexts, without explicitly inheriting validated rationales across iterations. As a result, they lack a traceable lineage and the produced factors are harder to audit and trust. \textit{(iii) Constrained exploration:} Search often over-exploits local neighborhoods around initial seeds, leading to redundancy and factor crowding, while providing limited systematic coverage of the broader hypothesis space and weakening long-horizon discovery.

To address these challenges, we propose \method, an evolutionary alpha mining framework that improves factor quality through trajectory-level self-evolution (Figure~\ref{fig:intro_combined}). Inspired by \cite{lin2025se}, we view each end-to-end mining run as a mining trajectory and explicitly evolve trajectories rather than relying on unconstrained re-generation from noisy feedback. First, to mitigate local optimization, we introduce a Diversified Planning Initialization that generates multiple complementary research directions, yielding broad initial coverage in the hypothesis space.  {Second}, to support controllable refinement and reliable reuse of validated experience, \method{} performs self-evolution via trajectory-level mutation and crossover. Mutation performs targeted revision by localizing the failure-causing step via self-reflection and rewriting only the corresponding segment, while keeping the rest of the trajectory intact. Crossover recombines complementary segments from high-reward parents to reuse validated hypothesis structures, construction patterns, and repair behaviors. We further apply generation gates to enforce semantic consistency and constrain complexity and redundancy, preventing drift and crowding. Together, these mechanisms expand exploration while stabilizing refinement under real-market noise and non-stationarity. Extensive experiments on CSI 300 and cross-market transfer to CSI 500 and S\&P 500 demonstrate consistent improvements in predictive power and strategy performance over strong baselines and prior agentic systems.



%% file: main_section/2.related_work.tex
\section{Related Work}

\textbf{Agents in Finance.} 
Recent advances in financial LLMs \citep{liu2023fingpt, liu2025fin} and evaluation benchmarks \citep{guo2025fineval, guo2026bizfinbench, tang2025finmmr} have substantially expanded the scope of automated financial reasoning.
Building on these foundations, a growing line of work explores how agentic systems \citep{li2025investorbench, yang2026finvault} can move beyond financial understanding toward {decision-oriented research workflows}, including factor discovery and trading analysis.
In this context, the pursuit of alpha factors has evolved from manual engineering and heuristic search toward LLM-driven closed-loop discovery \cite{chen2025alphasage, li2024can, shi2025navigating, wang2025alpha}.
Beyond search optimization, recent efforts emphasize workflow systematization. RD-Agent \cite{li2025r} proposes a multi-agent framework that decouples the pipeline into research and development stages, enabling data-centric joint optimization of factors and models. To address market non-stationarity, AlphaForge \cite{shi2025alphaforge} focuses on the dynamic combination of mined factors, 
while Alphafin \cite{li2024alphafin} and AlphaEval \cite{ding2025alphaeval} establish standardized, task-oriented protocols for reproducible evaluation. Despite these advances, trading constraints (e.g., turnover, complexity) remain largely post-hoc filters rather than intrinsic objectives, leading to suboptimal generalization and interpretability in live trading.

\textbf{Self-Evolving Agents.} Self-evolving agents \cite{fang2025comprehensive, lin2025se, zhai2025agentevolver, zhang2026evofsm} represent a paradigm shift from static instruction-following to autonomous learning through interacting with environment. 
AlphaEvolve \cite{novikov2025alphaevolve} demonstrates a coding-centric evolution where the agent employs evolutionary resampling to autonomously generate algorithms for scientific discovery. Building on this, CSE \cite{hu2026controlled} introduces controllable self-evolution, facilitating a critical transition from stochastic generation to feedback-driven evolution, while CogAlpha \cite{liu2025cognitive}  applies code-level evolutionary search to alpha mining.
The financial domain adapts this self-evolutionary framework to handle non-stationary and high-dimensional data. Research has converged on multi-agent coordination and structured feedback to guide evolution \cite{li2025hedgeagents}. TradingAgents \cite{xiao2024tradingagents} and FactorMAD \cite{duan2025factormad} utilize institutional-style debates to refine trading hypotheses, while QuantAgents \cite{li2025quantagents} and ATLAS \cite{papadakis2025atlas} incorporate simulated trading performance as a reward signal for dynamic prompt optimization. To ensure consistency over long horizons, FinMem \cite{yu2025finmem} and FinCon \cite{yu2024fincon} leverage hierarchical memory and conceptual reinforcement to retain and refine high-level trading experience.
Despite these advances, applying self-evolving agents to finance is hindered by low signal-to-noise ratios and delayed market feedback. To mitigate this, \method{} employs stable optimization units with mutation and crossover operators, ensuring a robust, traceable evolution path through structured archiving.

%% file: main_section/3.problem_formulation.tex
\section{Problem Setup}

\textbf{Alpha Mining.}
The alpha mining task aims to learn an alpha factor $f$ from a market feature tensor $\mathbf{X}\in\mathbb{R}^{N\times T\times D}$ for a stock universe $\mathcal{S}=\{s_1,\ldots,s_N\}$ and a time horizon $\mathcal{T}=\{t_1,\ldots,t_T\}$. 
At each time $t\in\mathcal{T}$, the factor produces a signal from the feature slice $\mathbf{X}_t\in\mathbb{R}^{N\times D}$ that is used to predict the cross-sectional return $\mathbf{y}_{t+1}\in\mathbb{R}^{N}$.
In notation, an alpha can be written as $f(\mathbf{X}_t)\rightarrow \mathbf{y}_{t+1}$, where $\mathbf{y}_{t+1}$ denotes the realized returns at time $t+1$. Accordingly, we formulate alpha mining as the following optimization:
\begin{equation}
f^{*}=\arg\max_{f\in\mathcal{F}}\ \mathcal{L}\big(f(\mathbf{X}),\mathbf{y}\big)-\lambda\mathcal{R}(f),
\label{eq:alpha_mining}
\end{equation}
where $\mathcal{F}$ denotes the space of all possible factor expressions, $\mathbf{y}$ is the ground-truth return target (e.g., next-day returns), $\mathcal{L}(\cdot)$ measures predictive effectiveness, $\mathcal{R}(\cdot)$ is a regularization term that encourages simplicity and novelty of the expression, and $\lambda$ balances utility and regularization.

\textbf{Alpha Mining Trajectory.}
Distinct from purely data-driven pipelines, we introduce market hypotheses $h\in\mathcal{H}$ to guide LLM-based factor construction. 
Each alpha mining run follows an end-to-end workflow from hypothesis generation to factor generation and backtesting evaluation (see Section~\ref{sec:4.1}). 
We represent each run by a mining trajectory, defined as an ordered sequence $\tau=(s_0,a_0,s_1,a_1,\ldots,s_n)$, where $s_0$ denotes the initial mining context (e.g., market context and optional user-provided seeds), $a_i$ is the action taken at step $i$ by the multi-agent system, and $s_n$ is the terminal state containing the evaluated result of this run. The quality of a trajectory is measured by its terminal reward:
\begin{equation}
R(\tau)=\mathcal{L}\big(f_{\tau}(\mathbf{X}),\mathbf{y}\big)-\lambda\mathcal{R}(f_{\tau}),
\label{eq:traj_reward}
\end{equation}
where $f_{\tau}$ denotes the factor produced by trajectory $\tau$.

\textbf{Objective.} Our objective is to learn a trajectory generation policy $\pi$ for the multi-agent system that maximizes the expected terminal reward of the generated mining trajectory: 
\begin{equation}
\pi^{*}=\arg\max_{\pi}\ \mathbb{E}_{\tau\sim \pi}\bigl[R(\tau)\bigr],
\label{eq:objective_policy}
\end{equation}
where $\tau\sim\pi$ denotes the trajectory induced by repeatedly applying $\pi$ from the initial state $s_0$.

%% file: main_section/4.method.tex
\section{Method}
\begin{figure*}[htbp]
    \centering
    \includegraphics[width=0.9\textwidth]{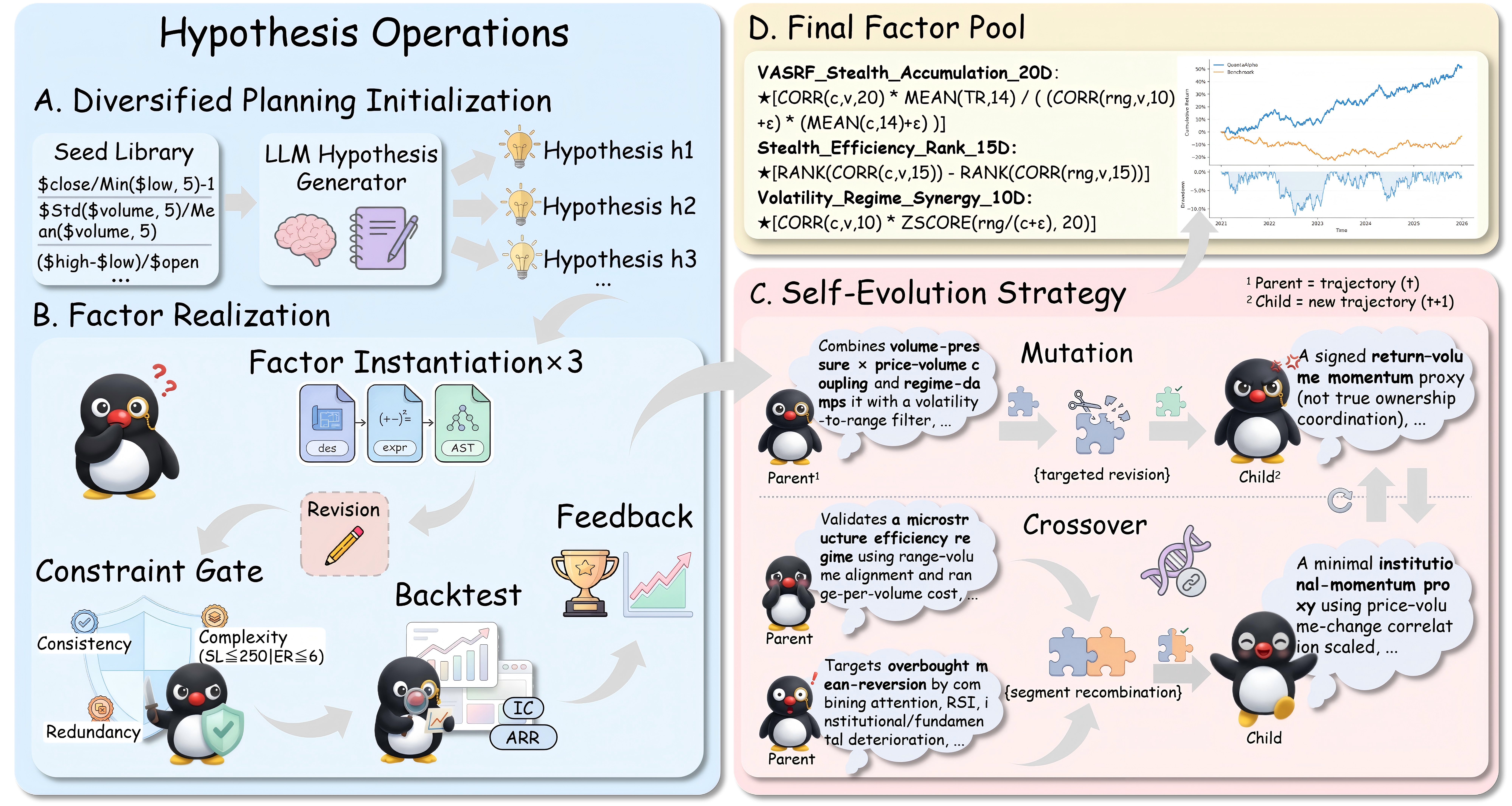}
    \caption{Overview of the QuantaAlpha framework, which consists of four core components: (A) Diversified Planning Initialization for candidate hypotheses, (B) Factor Realization that instantiates hypotheses into executable factors with constraint gating, (C) Self-Evolution with mutation and crossover over evaluated trajectories, and (D) A Final Factor Pool that consolidates validated factors.}
    \label{fig:overview}
\end{figure*}

As illustrated in Figure~\ref{fig:overview}, our approach treats alpha mining as an agentic workflow rather than one-shot factor construction. Given market context and user-provided seeds, a multi-agent system produces a mining trajectory including hypothesis generation, factor construction, implementation, and backtesting evaluation. We first describe the trajectory generation workflow (Section~\ref{sec:4.1}), and then introduce a self-evolution scheme that iteratively improves mining quality (Section~\ref{sec:4.2}).

\subsection{Alpha Mining Process} 
\label{sec:4.1}

A standalone alpha mining process progresses from hypothesis generation to factor creation and backtesting-based evaluation. We translate this process into collaboration among multiple agents, each responsible for a specific task at different stages of the workflow.

\subsubsection{Hypothesis Generation}
We use the idea agent $\mathcal{A}{i}$ to generate market hypotheses via structured knowledge integration. Conditioned on (i) observations from current market conditions or previous mining trajectories, (ii) domain priors from financial theories and empirical evidence, and (iii) parametric specifications (e.g., ``10-day high/low''), $\mathcal{A}{i}$ produces actionable hypotheses that describe candidate market mechanisms and serve as the input to subsequent factor generation.



\subsubsection{Controllable Factor Construction}
To effectively instantiate hypothesis-driven factors, we need a robust mechanism to align implementations with domain hypotheses. However, direct code generation is brittle, often suffering from syntax errors, dependency mismatches, and semantic drift in complex implementations, which undermines fidelity to the hypothesis. To address these limitations, We introduce an intermediate symbolic representation based on a standardized operator library $\mathcal{O}$ and an Abstract Syntax Tree (AST). This abstraction bridges high-level market intent and low-level implementation, enabling controllable construction, structural validation, and reliable compilation.

\textbf{Factor Generation.} Given a hypothesis $h \in \mathcal{H}$ and raw features $\mathcal{X}$, the factor agent $\mathcal{A}_{f}$ generates a symbolic expression $f \in \mathcal{F}$ over the operator library and parses it into an AST as the intermediate representation. Concretely, it first maps $h$ to a structured semantic description $d$ that formalizes the intended mechanism into discrete operators, while concurrently instantiating operational parameters—such as look-back windows and thresholds—from either prescribed parameters in $h$ or heuristic defaults anchored in domain expertise.
Conditioned on $d$, the agent then assembles an operator composition into a symbolic expression $f$ and parses it into an AST, denoted as $T(f)$. In $T(f)$, leaf nodes bind to raw feature fields (e.g., \texttt{\$high}, \texttt{\$volume}) and internal nodes correspond to operator instances from $\mathcal{O}$ (e.g., \texttt{TS\_MIN}($\cdot$), \texttt{SMA}($\cdot$), \texttt{RANK}$(\cdot)$), thereby rendering the computational dependencies and data flow fully transparent.
Finally, the agent translates the validated symbolic form into executable code $c$ via compilation; when compilation fails due to implementation issues, an LLM-based repair step is triggered to regenerate a consistent implementation while preserving the semantics of $f$. This design ensures that code generation remains anchored to an explicit symbolic specification rather than unconstrained free-form generation.

\textbf{Consistency Verification.} We enforce semantic consistency across representations to prevent drift during generation. Specifically, we apply an LLM-based verifier to assess (i) alignment among the hypothesis $h$, semantic description $d$, and symbolic expression $f$, and (ii) faithfulness between the symbolic definition $f$ and the generated code $c$. The verifier returns a consistency decision under fixed criteria. If it fails, we rewrite the inconsistent component(s) by regenerating $d$ and $f$ or repairing $c$ until the check passes or a maximum retry budget is reached.

\textbf{Complexity and Redundancy Control.} We further regularize factor generation with explicit structural constraints to promote parsimony and novelty. For complexity, we compute
\begin{equation}
\mathcal{C}(f)
= \alpha_1 \cdot SL(f)
+ \alpha_2 \cdot PC(f)
+ \alpha_3 \cdot \log\!\bigl(1 + |F_f|\bigr),
\label{eq:factor_complexity}
\end{equation}
where $SL(f)$ measures symbolic length, $PC(f)$ counts free parameters (e.g., window sizes), and $F_f$ is the set of features used by $f$.
For redundancy, we quantify structural similarity via AST matching. Given two factors $f_i$ and $f_j$ with ASTs $T(f_i)$ and $T(f_j)$, we define $s(f_i,f_j)$ as the size of the largest common subtree:
\begin{equation}
s(f_i,f_j)
=\max_{\substack{S_i \subseteq T(f_i), S_j \subseteq T(f_j), S_i \cong S_j}} |V(S_i)|.
\end{equation}
Here, $S_i$ and $S_j$ range over subtrees of $T(f_i)$ and $T(f_j)$, $S_i \cong S_j$ denotes subtree isomorphism, and $|V(S_i)|$ is the number of nodes in $S_i$. For a candidate factor $f$, we compute its maximum similarity to an existing alpha zoo $\mathcal{Z}$ as
$
S(f)=\max_{\phi\in\mathcal{Z}} s(f,\phi).$

We reject and rewrite any factor that violates the prescribed complexity or redundancy limits, promoting parsimonious generation while avoiding near-duplicate structures. We further apply output-level correlation filtering on the 2021 validation panel to remove functionally equivalent factors: if two factors exceed an absolute correlation threshold, the one with lower RankIC is discarded. This two-stage design handles both structural and functional redundancy.

\subsubsection{Factor Evaluation} 

The evaluation agent assesses factors via a standardized backtesting system. The protocol covers three aspects: predictive capability metrics for forecasting effectiveness, return performance metrics for profit-generating potential under a fixed execution setting, and risk control metrics for stability and robustness across conditions. Beyond factor scoring, the agent maintains an evaluation history that records outcomes and summarizes systematic patterns among successful and failed factors.

\subsection{Evolutionary Alpha Mining}
\label{sec:4.2}

We improve alpha discovery via iterative optimization over mining trajectories. Starting from user-provided seed factors, we generate a diversified set of initial hypotheses. We then run the workflow in Section~\ref{sec:4.1} to turn each hypothesis into an evaluated mining trajectory, forming an initial trajectory pool $\mathcal{T}_0=\{\tau_j^0\}_{j=1}^{N_{init}}$. We then apply an evolution process to obtain progressively improved factors.

\subsubsection{Diversified Planning Initialization}
\label{sec:4.2.1}
Let $\mathcal{Z}_0$ denote the user-provided seed factors (or seed ideas). The initial seed pool is derived from the public Alpha158(20) subset, grouped by low correlation to ensure diverse starting directions. 
We employ an initialization agent $\mathcal{A}_0$ to propose a diversified set of initial hypotheses $\mathcal{H}^{0}=\{h^{0}_1,\ldots,h^{0}_{N_{init}}\}$. The agent is instructed to maximize complementarity among hypotheses at both semantic and structural levels, e.g., varying signal sources (price vs.\ volume), time scales (short-term vs.\ long-term), and mechanism types (momentum vs.\ mean-reversion or regime-conditioned triggers). For each $h^{0}_i \in \mathcal{H}^{0}$, we run the mining workflow in Section~\ref{sec:4.1} and obtain a mining trajectory $\tau^{0}_i$.

This diversified initialization expands the effective search frontier by providing broad coverage in the hypothesis space. As a result, the subsequent evolutionary process can explore multiple promising regions in parallel, reducing the risk of premature convergence to a suboptimal local optimum.

\subsubsection{Self-Evolution}
\label{sec:4.2.2}
The goal of self-evolutionary updates is to guide the alpha mining system to search for better trajectories based on existing ones. Specifically, we apply trajectory-level operators to existing mining trajectories to generate improved trajectories as demonstrations. These demonstrations provide imitation learning priors that bias subsequent trajectory generation toward effective decisions. In practice, we instantiate the update with two evolutionary primitives, Mutation and Crossover, which revise a single trajectory or recombine complementary sub-trajectories, respectively. Each iteration $i$ yields a new trajectory generation $\mathcal{T}_i$, and factor quality improves progressively across iterations (see Figure~\ref{fig:case_study} in Appendix for a case study).

\textbf{Mutation.}
Given a mining trajectory $\tau \in \mathcal{T}_{i-1}$, the agent first performs self-reflection to diagnose sub-optimal decision node that most significantly accounts for the low terminal reward,
denoted by an index $k\in\{0,\ldots,n-1\}$. 
We then rewrite only the localized action (or a short local segment) while keeping the remaining steps unchanged:
\begin{equation}
\tau_{child}=\big(s_0,a_0,\ldots,s_k,\mathrm{Refine}(a_k),s_{k+1}',a_{k+1}',\ldots,s_n'\big), 
\label{eq:controlled_mutation}
\end{equation}
where the prefix up to $s_k$ is frozen, and the remaining steps are regenerated by the agent conditioned on this fixed prefix to preserve trajectory coherence. Depending on the localized fault, the rewrite may update the hypothesis, the symbolic expression under the operator library, or the compiled code, and can introduce mechanism-level changes such as altering the time scale, or adding regime conditions. The diagnostic step provides a directional search signal, allowing even imperfect localization to move the child trajectory toward a different region of the factor space.

\textbf{Crossover.}
Crossover aims to exploit validated components by recombining complementary sub-trajectories from high-quality parents. At iteration $i-1$, we select a parent subset $\mathcal{P}_{i-1}\subseteq \mathcal{T}_{i-1}$ based on trajectory reward (Eq.~\ref{eq:traj_reward}). Given $k$ parent trajectories $\tau^{(1)},\ldots,\tau^{(k)}\in\mathcal{P}_{i-1}$, Crossover synthesizes a child trajectory by composing high-performing segments from different parents:
\begin{equation}
\tau_{\mathrm{child}} = \mathrm{Crossover}\big(\tau^{(1)},\ldots,\tau^{(k)}\big).
\label{eq:crossover}
\end{equation}
Operationally, the primitive identifies trajectory segments that consistently contribute to high cumulative rewards—such as hypothesis templates, factor construction patterns, or strategic repair actions—and merges them into a highly coherent sequence. This recombination mimics the human practice of combining complementary strengths from different solutions, while providing a more credible lineage by explicitly inheriting decisions validated in previously successful trajectories.

%% file: main_section/5.experiments.tex
\section{Experiments}

\subsection{Experimental Setup}

\textbf{Datasets and Metrics.} We conduct experiments on the CSI 300 dataset, which covers 300 large-cap A-share stocks in the Chinese market. We adopt a chronological split with training (Jan. 1, 2016 to Dec. 31, 2020), validation (Jan. 1, 2021 to Dec. 31, 2021), and testing (Jan. 1, 2022 to Dec. 26, 2025). We evaluate performance from two complementary perspectives. Factor predictive power is measured by Information Coefficient (IC), IC Information Ratio (ICIR), Rank IC, and Rank ICIR. Strategy performance is measured by Annualized Return (ARR), Information Ratio (IR), and Maximum Drawdown (MDD). Further details are provided in Appendix~\ref{appendix:A.2}.

\textbf{Baselines.} We compare QuantaAlpha against four baseline categories: traditional machine learning models, deep learning time-series models, classical factor libraries, and LLM-based factor research agents, including RD-Agent and AlphaAgent. More details are provided in Appendix~\ref{appendix:A.5}. For all LLM-based methods, roughly 150 validated factors are submitted to the same downstream LightGBM model for final evaluation, ensuring a fair factor-pool-level comparison.






\subsection{Main Results}
Table~\ref{tab:csi300_results} reports the main results on the CSI 300 market over a four-year evaluation period, and strategy-level results are computed from a factor pool of approximately 150 validated factors synthesized by the same downstream LightGBM model, rather than from a single best factor. For transparency regarding the search budget, we report computational cost and token consumption in Appendix~\ref{sec:computational_cost}. QuantaAlpha achieves the strongest overall performance across most factor predictive power and strategy-level performance. Using GPT-5.2, QuantaAlpha achieves the best IC at 0.0472, delivers an ARR of 4.68\%, and maintains a low MDD of 11.80\%. From an empirical perspective, these results suggest that the mined factors capture non-trivial predictive signals beyond simple backtest noise, although real-world deployment would require additional transaction-cost modeling, risk control, and live trading validation. We also observe broadly consistent gains across different base models, suggesting that the improvements are not tied to a single LLM backbone. The narrower gap between Claude-4.5-Sonnet and DeepSeek-V3.2 within QuantaAlpha (0.004 IC) is attributed to backbone-specific adaptation: Claude's longer feedback outputs reduce the number of prior trajectories fitting within the fixed context budget, and its mutations tend to be more local, whereas DeepSeek-V3.2 more frequently proposes larger exploratory jumps.

Compared with RD-Agent, QuantaAlpha and AlphaAgent both incorporate generation-stage regularization, and this is reflected in stronger predictive power and strategy performance. Under GPT-5.2, QuantaAlpha improves IC by 0.0186 and ARR by 1.10\% relative to RD-Agent while reducing MDD by 4.96\%, suggesting that constraining hypothesis-to-factor construction with explicit intermediate representations and verification gates effectively reduces semantic drift and improves implementation reliability. Beyond these constraints, QuantaAlpha further improves upon AlphaAgent by 0.0125 IC and 3.57\% ARR while reducing MDD by 2.09\%, highlighting the added value of trajectory-centric self-evolution: mutation broadens exploration via mechanism-level variations, while crossover reuses validated trajectory segments and repair strategies, jointly increasing the yield and stability of high-quality factors under non-stationary market dynamics.


\input{tables/main_table}



\subsection{Ablation Study}
\begin{figure*}[htbp]
\centering
\begin{minipage}{0.4\textwidth}
  \centering
  \small
  \captionof{table}{Ablation study of evolutionary mining components.}
  \label{tab:ablation_study_two_column}
  \resizebox{\linewidth}{!}{%
  \begin{tabular}{lcccc}
    \toprule
    \multirow{2}{*}{Method} & \multicolumn{4}{c}{Key Metrics} \\
    \cmidrule(lr){2-5}
     & \multicolumn{1}{l}{\,\,\,\textbf{IC}} 
     & \multicolumn{1}{l}{\textbf{Rank IC}} 
     & \multicolumn{1}{l}{\textbf{ARR} (\%)} 
     & \textbf{MDD} (\%) $\downarrow$\\
    \midrule

    \textbf{QuantaAlpha}
    & \multicolumn{1}{l}{\textbf{0.0461}}
    & \multicolumn{1}{l}{\textbf{0.045}}
    & \multicolumn{1}{l}{\textbf{4.53}}
    & \multicolumn{1}{l}{\,\,\,\,\textbf{15.1}} \\

    \phantom{Q} - \emph{w/o} \textbf{Planning}
    & 0.0448\textsuperscript{\textcolor[RGB]{230,180,180}{\tiny -0.0013}}
    & 0.0437\textsuperscript{\textcolor[RGB]{230,180,180}{\tiny -0.0013}}
    & 3.81\textsuperscript{\textcolor[RGB]{180,20,20}{\tiny -0.72}}
    & 16.72\textsuperscript{\textcolor[RGB]{200,50,50}{\tiny +1.62}} \\

    \phantom{Q} - \emph{w/o} \textbf{Mutation}
    & 0.0382\textsuperscript{\textcolor[RGB]{180,60,60}{\tiny -0.0079}}
    & 0.0371\textsuperscript{\textcolor[RGB]{180,60,60}{\tiny -0.0079}}
    & 3.27\textsuperscript{\textcolor[RGB]{160,20,20}{\tiny -1.26}}
    & 15.58\textsuperscript{\textcolor[RGB]{220,160,160}{\tiny +0.48}} \\

    \phantom{Q} - \emph{w/o} \textbf{Crossover}
    & 0.0401\textsuperscript{\textcolor[RGB]{220,150,150}{\tiny -0.0060}}
    & 0.0419\textsuperscript{\textcolor[RGB]{220,150,150}{\tiny -0.0031}}
    & 4.02\textsuperscript{\textcolor[RGB]{200,50,50}{\tiny -0.51}}
    & 16.03\textsuperscript{\textcolor[RGB]{220,100,100}{\tiny +0.93}} \\

    \bottomrule
  \end{tabular}
  }
\end{minipage}
\hfill
\begin{minipage}{0.58\textwidth}
  \centering
  \includegraphics[width=\linewidth]{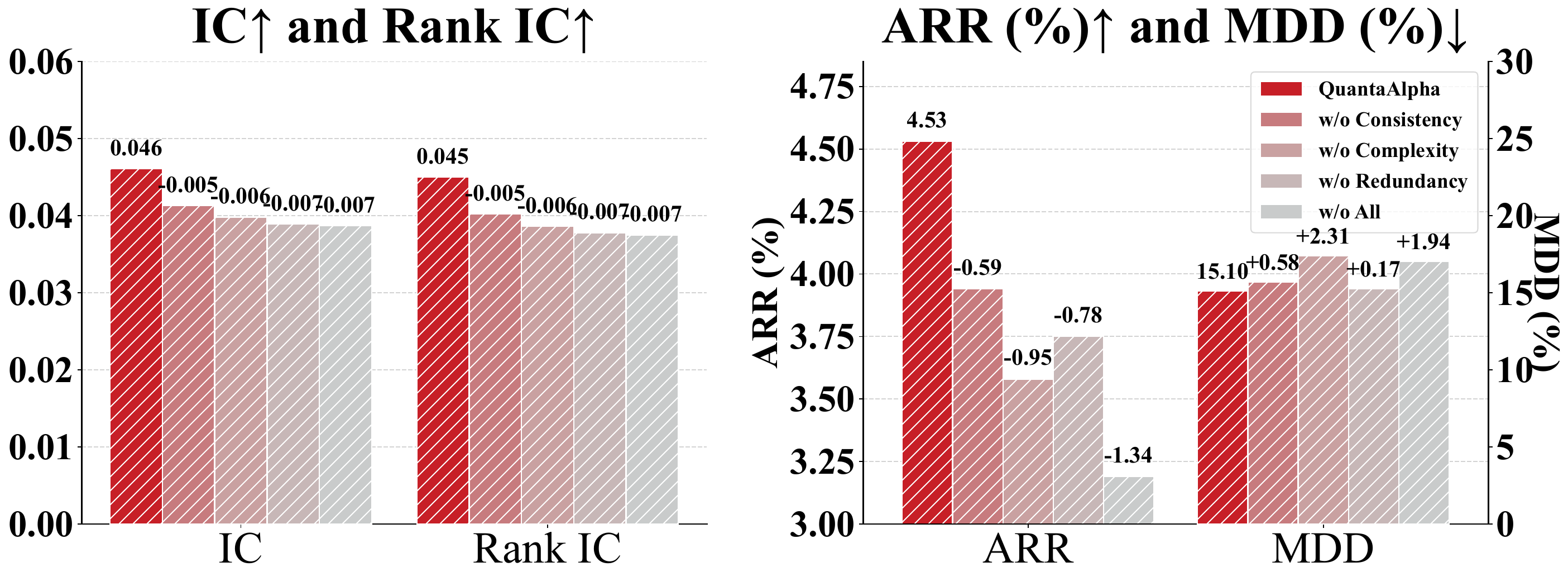}
  \caption{Ablation study of semantic consistency, complexity, and redundancy controls.}
  \label{fig:gate_ablation}
\end{minipage}
\end{figure*}

\textbf{Ablation of Evolutionary Mining Components.} We conduct an ablation study to quantify the contribution of three components in evolutionary alpha mining: diversified planning initialization, trajectory mutation, and trajectory crossover. The results are summarized in Table~\ref{tab:ablation_study_two_column}. Removing \emph{planning} yields only marginal changes in IC and Rank IC, but substantially degrades strategy-level outcomes, with ARR decreasing by 0.72\% and MDD increasing by 1.62\%. This indicates that diversified initialization primarily improves the search frontier by providing broader and less correlated starting hypotheses, which stabilizes subsequent evolution. In contrast, removing \emph{mutation} causes the largest drop in predictive power, decreasing IC by 0.0079 and Rank IC by 0.0079, and also leads to the largest reduction in ARR by 1.26\%. This highlights mutation as the primary driver of effective exploration and trajectory repair, enabling the system to escape suboptimal regions and correct failure modes discovered during mining. Meanwhile, removing \emph{crossover} leads to a smaller but consistent degradation. This supports the role of crossover in exploiting and inheriting complementary high-performing trajectory segments, improving efficiency and stability by recombining validated patterns from successful trajectories. These results suggest that mutation contributes more to exploration in the current five-cycle setting, whereas crossover mainly supports reuse and recombination of validated trajectory segments. More broadly, QuantaAlpha's contribution lies in evolving complete research trajectories rather than relying on any single operator.

\textbf{Ablation of Consistency, Complexity, and Redundancy Controls.} We ablate three controls during factor generation that gate rewriting, including semantic consistency verification, complexity regularization, and redundancy filtering. As shown in Figure~\ref{fig:gate_ablation}, disabling any single control consistently degrades performance, confirming that each contributes non-trivially to robust factor generation. The \emph{consistency} gate prevents semantic drift between the hypothesis, symbolic specification, and implementation. The \emph{complexity} control improves robustness by discouraging overly complex expressions that generalize poorly, and its removal leads to the most pronounced degradation at the strategy level, with the annualized excess return dropping by 0.95\% and the maximum drawdown increasing by 2.31\%. The \emph{redundancy} control preserves exploration by filtering near-duplicate structures and mitigating factor crowding. Disabling all three yields the largest degradation, indicating that these controls are complementary and jointly needed for reliable generation.

We further analyze cross-seed variance and evolutionary robustness in Appendix~\ref{sec:cross_seed_robustness}, showing that QuantaAlpha remains stable across different initial seed combinations and that the evolutionary process does not drift uncontrollably.

\subsection{More Analysis}

\textbf{Factor Generalizability.} To evaluate factor robustness under significant market distribution shifts, we conduct a zero-shot transfer experiment where the factors mined and selected on CSI 300 are directly deployed on CSI 500 and the S\&P 500 without any re-optimization or market-specific adaptation. Zero-shot transfer is feasible because the factors use standard OHLCV-based operators and daily cross-sectional normalization mitigates cross-market scale differences. As shown in Figure~\ref{fig:intro_combined}, the factors mined on CSI 300 exhibit positive transfer performance on both CSI 500 and the S\&P 500. On CSI 500, QuantaAlpha remains broadly comparable to other alpha-mining baselines during the early evaluation period, but shows a stronger recovery and a higher terminal cumulative return after late 2024, outperforming AlphaAgent, RD-Agent, Alpha158, and the benchmark by the end of the test period. On the S\&P 500, QuantaAlpha also achieves competitive out-of-sample performance and attains the highest terminal return.  By the end of the evaluation period, it reaches a cumulative excess return of roughly \textbf{40.28\%} on CSI 500 and over \textbf{19.1\%} on the S\&P 500. These results indicate that the discovered factors transfer beyond the source market and retain effectiveness under distribution shifts, rather than relying on market-specific historical patterns.



\begin{figure*}[htbp]
\centering

\begin{minipage}{0.62\textwidth}
\centering
\includegraphics[width=\linewidth]{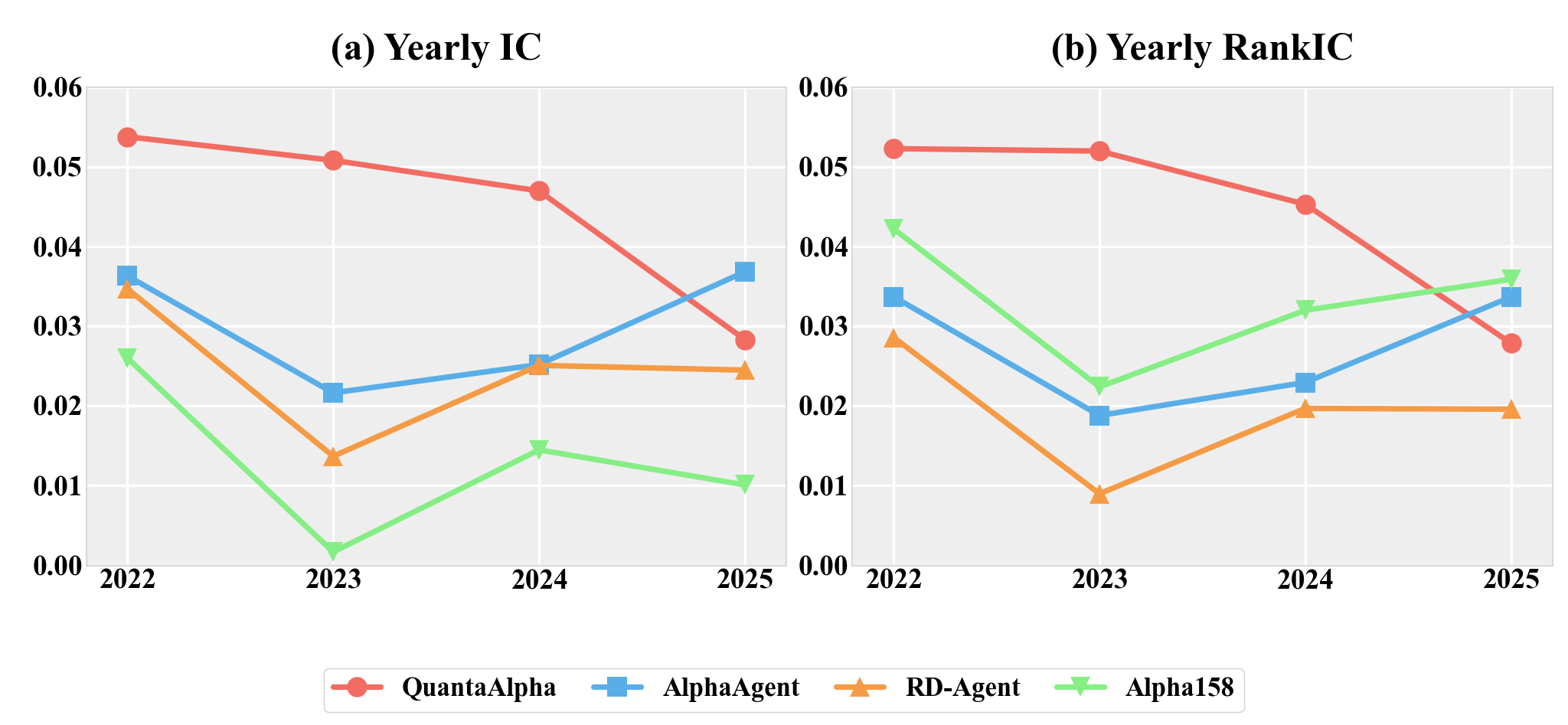}
\caption{Annual IC and Rank IC comparison on the CSI 300.}
\label{fig:annual_ic_rankic}
\end{minipage}
\hfill
\begin{minipage}{0.36\textwidth}
\centering
\includegraphics[width=\linewidth]{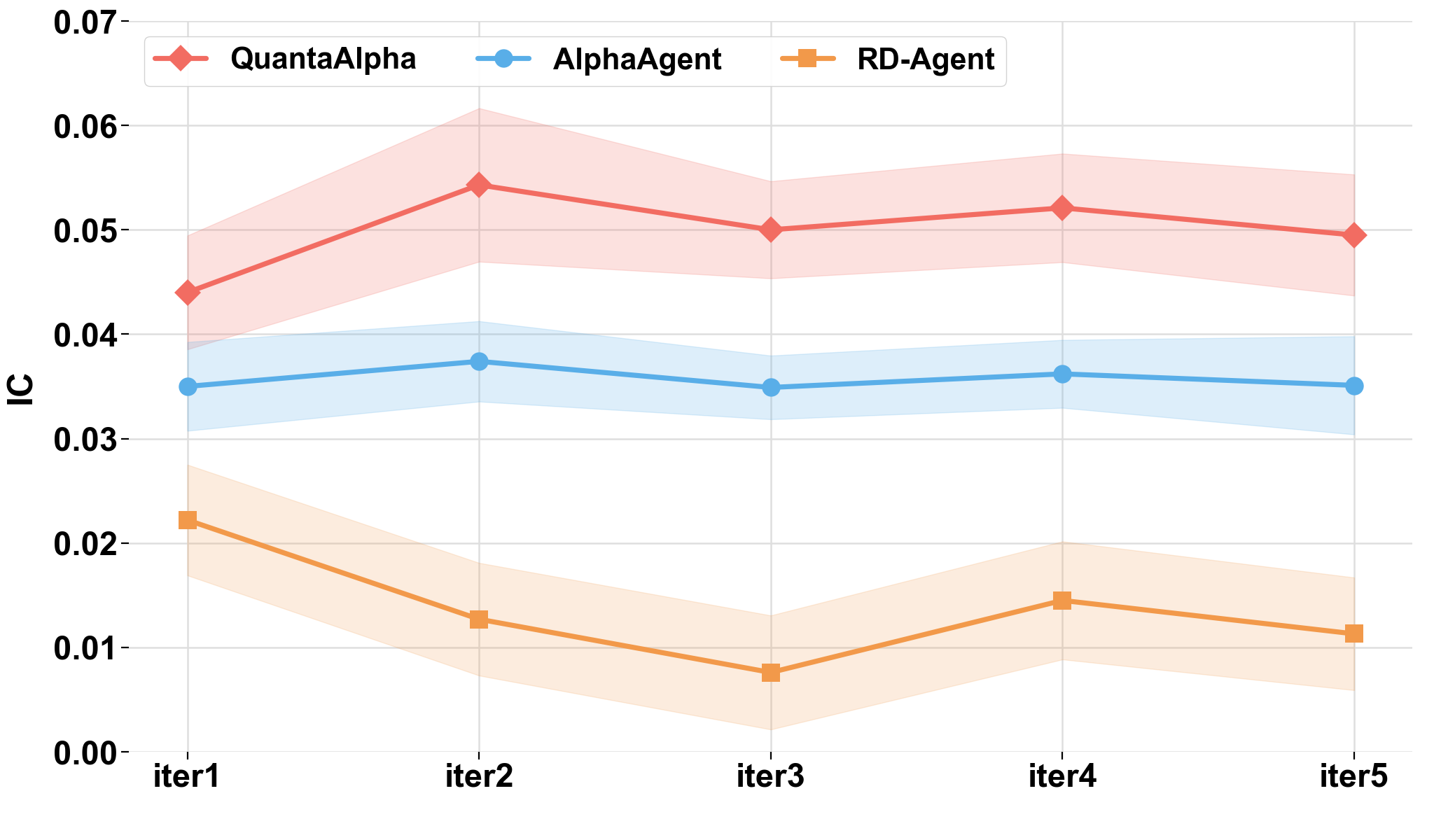}
\caption{The evolution of IC over the first five iterations.}
\label{fig:ic_evolution}
\end{minipage}

\end{figure*}

\textbf{Alpha Decay.} Figure~\ref{fig:annual_ic_rankic} reports annual IC and Rank IC on the CSI~300 universe from 2021 to 2025. A pronounced performance collapse is observed for the baselines in 2023, coinciding with a major regime shift in the A-share market. The pre-2023 period is dominated by large-cap ``core assets'', where institutional trading supports stable trends and makes classical momentum or mean-reversion signals effective. In 2023, the market rotates toward small-cap and thematic stocks, with higher intraday noise, more frequent overnight gaps, and weaker trend persistence. Baseline methods, whose factor libraries assume smooth trends and regular reversal patterns, fail to transfer to this out-of-distribution environment. QuantaAlpha, by comparison, maintains strong IC and Rank IC through the regime transition. It discovers structural factors, such as \textit{Mean-Reverting Range Deviation} and \textit{Overnight Gap Structure}, that reflect persistent microstructure effects including volatility clustering and overnight information incorporation. These signals are less sensitive to capitalization style, enabling better temporal generalization. A factor-level analysis is provided in Appendix~\ref{app:factor_analysis}.

\textbf{Evolutionary Alpha Mining Efficiency.} We analyze the iterative dynamics of factor mining by tracking the IC distribution across generation rounds. As visualized in Figure~\ref{fig:ic_evolution}, QuantaAlpha consistently maintains the highest IC across all five iterations, indicating that its evolutionary updates improve factor quality more effectively than both AlphaAgent and RD-Agent. Notably, QuantaAlpha exhibits a rapid gain in the early rounds and then remains stable at a high level, suggesting strong sample efficiency in converting early exploration into consistently predictive factors. In addition to the higher mean, QuantaAlpha shows a moderate but persistent spread of IC across rounds, reflecting sustained diversity in the explored factor candidates rather than premature collapse to a narrow region.  By comparison, RD-Agent remains substantially lower across all iterations, suggesting less effective iterative refinement and weaker accumulation of useful generation experience.  AlphaAgent improves over RD-Agent but remains consistently below QuantaAlpha, suggesting that our trajectory-level evolution more efficiently accumulates and reuses successful generation patterns across iterations.


\subsection{Case study}
To make the evolution process concrete, we present a representative case study in Figure~\ref{fig:case_study} of the Appendix and trace how QuantaAlpha updates hypotheses and factors over five iterations. The first iteration yields interpretable short-term reversal factors. The second broadens the mechanism via volatility-weighted momentum, but increased structural complexity coincides with weaker generalization. Subsequent iterations simplify the expression into a linear additive form, improving drawdown and stabilizing performance. The fifth iteration adds participant-differentiated behavioral signals, incorporating complementary information and further improving predictability. Throughout the evolution, QuantaAlpha maintains a cumulative factor pool. Its predictive performance does not saturate within the first five iterations and only begins to decay after roughly 15 iterations, indicating that the evolution sustains effective improvement over a substantial horizon before diminishing returns emerge. Detailed analysis is provided in Appendix~\ref{sec:appendix_iteration_convergence}.



%% file: tables/main_table.tex
\begin{table*}[ht]
\centering
\caption{\footnotesize
Performance comparison of different methods on \textit{\textbf{CSI 300}}.
We report both \textit{factor predictive power} and \textit{strategy-level performance} metrics, where \textbf{higher$\uparrow$} values indicate better performance for all metrics, except for MDD where lower values are better.
For each method category, the best result in each column is highlighted in \textbf{bold},
while the second-best result is \underline{underlined}.
Across all methods, the overall best-performing method is highlighted with a light \colorbox[RGB]{236,244,252}{blue background}.
}
\label{tab:csi300_results}
\setlength\tabcolsep{3.2pt}

\fontsize{8.1pt}{10.5pt}\selectfont

\resizebox{1\textwidth}{!}{%

\begin{tabular}{llccccccc}

\toprule[1.5pt]

\multicolumn{2}{c}{\multirow{2}{*}{\textbf{Methods}}} 

& \multicolumn{4}{c}{Factor Predictive Power}

& \multicolumn{3}{c}{Strategy Performance} \\

\cmidrule(lr){3-6} \cmidrule(lr){7-9}

\multicolumn{2}{c}{} 

& \textbf{IC} & \textbf{ICIR} & \textbf{Rank IC} & \textbf{Rank ICIR} 

& \textbf{IR (SHR*)} & \textbf{ARR (\%)} & \textbf{MDD (\%)$\downarrow$} \\

\midrule[0.8pt]

\rowcolor[RGB]{230, 230, 230}\multicolumn{9}{c}{\textit{\textbf{Classical Factor Mining}}} \\

\multirow{6}{*}{\textbf{Machine Learning}}
& Linear 
& 0.0155 
& 0.1174 
& 0.0368 
& 0.2834 
& -0.3078 
& -2.67 
& 18.97 \\

& XGBoost 
& 0.0175 
& 0.1336 
& 0.0420 
& 0.3417 
& -0.5280 
& -4.24 
& 28.50 \\

& CatBoost 
& 0.0162 
& 0.1203 
& 0.0405 
& 0.3289 
& -0.2807 
& -2.30 
& 21.35 \\

& LightGBM 
& \underline{0.0247} 
& \underline{0.2055} 
& \underline{0.0423} 
& \underline{0.3726} 
& 0.0092 
& 0.07 
& 21.80 \\

& MLP 
& \textbf{0.0321} 
& \textbf{0.2780} 
& \textbf{0.0438} 
& \textbf{0.4088} 
& \underline{0.1716} 
& \underline{1.46} 
& \underline{18.15} \\

& DoubleEnsemble 
& 0.0213 
& 0.1670 
& 0.0408 
& 0.3372 
& \textbf{0.2490} 
& \textbf{1.85} 
& \textbf{15.00} \\

\midrule

\multirow{4}{*}{\textbf{Deep Learning}}
& GRU 
& 0.0321 
& 0.2603 
& 0.0442 
& 0.3601 
& 0.5302 
& 3.61 
& 15.01 \\

& Transformer 
& \underline{0.0331} 
& \underline{0.2702} 
& \underline{0.0451} 
& \underline{0.3801} 
& 0.4502 
& 5.21
& \underline{13.81} \\

& LSTM 
& \underline{0.0331} 
& 0.2502 
& \underline{0.0451} 
& 0.3503 
& \underline{0.6802} 
& \underline{6.01} 
& 14.81 \\

& TRA 
& \textbf{0.0421} 
& \textbf{0.3402} 
& \textbf{0.0511} 
& \textbf{0.4203} 
& \textbf{1.0502} 
& \textbf{6.81} 
& \textbf{8.51} \\

\midrule

\multirow{3}{*}{\textbf{Factor Libraries}}
& Alpha158(20) 
& 0.0051 
& 0.0329 
& 0.0184 
& 0.1177 
& \underline{0.5044} 
& \textbf{4.63} 
& 22.19 \\

& Alpha158 
& \textbf{0.0131} 
& \textbf{0.0817} 
& \textbf{0.0334} 
& \textbf{0.2119} 
& 0.4099 
& 2.66 
& \textbf{10.15} \\

& Alpha360 
& \underline{0.0105} 
& \underline{0.0636} 
& \underline{0.0306} 
& \underline{0.1889} 
& \textbf{0.6009} 
& \underline{4.09} 
& \underline{11.52} \\

\midrule[0.8pt]

\rowcolor[RGB]{230, 230, 230}\multicolumn{9}{c}{\textit{\textbf{LLM-based Agentic Factor Mining}}} \\

\multirow{5}{*}{\textbf{RD-Agent}}
& Qwen3-235B 
& 0.0267 
& 0.1676 
& 0.0194 
& 0.1199 
& -0.0818
& -0.62 
& 15.04 \\

& Deepseek-V3.2 
& 0.0245 
& 0.1630
& 0.0192 
& 0.1250 
& -0.2123
& -1.42 
& 19.17 \\

& Gemini-3-pro-preview 
& \textbf{0.0301} 
& 0.1870 
& \textbf{0.0282}
& 0.1677 
& 0.2595
& 1.89 
& \underline{11.49} \\

& Claude-4.5-sonnet 
& 0.0280 
& \textbf{0.2000} 
& 0.0242 
& \underline{0.1708} 
& \underline{0.3568} 
& \underline{2.36} 
& \textbf{10.81} \\

& GPT-5.2 
& \underline{0.0286} 
& \underline{0.1995} 
& \underline{0.0250} 
& \textbf{0.1739} 
& \textbf{0.5321} 
& \textbf{3.58} 
& 16.76 \\

\midrule

\multirow{5}{*}{\textbf{AlphaAgent}}
& Qwen3-235B 
& 0.0208 
& 0.1316 
& 0.0196 
& 0.1246 
& -0.0951 
& -0.60
& 18.56 \\

& Deepseek-V3.2 
& 0.0299 
& 0.1969 
& 0.0272 
& \underline{0.1799} 
& \underline{0.3972} 
& \underline{2.58} 
& \textbf{9.23} \\

& Gemini-3-pro-preview 
& 0.0263 
& 0.1671 
& 0.0236 
& 0.1512
& 0.1663
& 1.17 
& 14.05 \\

& Claude-4.5-sonnet 
& \underline{0.0311} 
& \underline{0.2043} 
& \underline{0.0286} 
& 0.1754 
& \textbf{0.4105} 
& \textbf{2.84} 
& 14.72 \\

& GPT-5.2 
& \textbf{0.0347} 
& \textbf{0.2122} 
& \textbf{0.0334} 
& \textbf{0.2053}
& 0.1587
& 1.11 
& \underline{13.89} \\

\midrule

\multirow{5}{*}{\colorbox[RGB]{245, 235, 235}{\textbf{QuantaAlpha}}}
& Qwen3-235B 
& 0.0450 
& 0.2538 
& 0.0444 
& 0.2507
& 0.3511
& 2.06
& 16.36 \\

& Deepseek-V3.2 
& \underline{0.0461} 
& \underline{0.2624} 
& \underline{0.0450} 
& \underline{0.2574}
& \underline{0.6271} 
& \underline{4.53} 
& 15.10 \\

& Gemini-3-pro-preview 
& 0.0453 
& 0.2551
& 0.0439
& 0.2490
& 0.5834 
& 4.21 
& \underline{12.10} \\

& Claude-4.5-sonnet 
& 0.0445 
& 0.2507 
& 0.0431 
& 0.2446 
& 0.5619
& 4.12 
& 13.02 \\

& GPT-5.2 
& \colorbox[RGB]{236,244,252}{\textbf{0.0472}}
& \colorbox[RGB]{236,244,252}{\textbf{0.2691}}
& \colorbox[RGB]{236,244,252}{\textbf{0.0459}} 
& \colorbox[RGB]{236,244,252}{\textbf{0.2635}}
& \colorbox[RGB]{236,244,252}{\textbf{0.6453}}
& \colorbox[RGB]{236,244,252}{\textbf{4.68}}
& \colorbox[RGB]{236,244,252}{\textbf{11.80}} \\

\bottomrule[1.5pt]
\end{tabular}}
\end{table*}


%% file: main_section/6.conclusion.tex
\section{Conclusion}
We present~\method, a self-evolving framework for interpretable alpha mining that formulates factor discovery as a constrained multi-agent process. Extensive experiments across Chinese and U.S. equity markets show that \method{} consistently produces more stable and generalizable factors than existing baselines. Beyond empirical performance, \method{} improves diversity through broad hypothesis exploration and redundancy-aware evolution, while maintaining controllability via symbolic representations and constraint-aware synthesis. These results suggest that agentic evolution is a promising paradigm for discovery problems in high-noise, non-stationary domains.

%% file: main_section/7.acknowledgement.tex
\section*{Impact Statement}


This paper presents work whose goal is to advance the field of Machine
Learning. There are many potential societal consequences of our work, none
which we feel must be specifically highlighted here.



%% file: appendix/1.app_1.tex

\section{Experiment Settings}
\label{sec:appendix_experiment_details}

This section provides details of our experimental setup, including computational infrastructure, evaluation metrics, backtesting setup, and baselines. Data and code are available at \url{https://github.com/QuantaAlpha/QuantaAlpha}.

\subsection{Evaluation Metrics}
\label{appendix:A.2}

We evaluate predictive performance using two categories of metrics: factor predictive power and strategy-level performance.

Without loss of generality, the bar notation $\bar{\cdot}$ denotes the mean, and $\sigma(\cdot)$ denotes the standard deviation.

\begin{itemize}
    \setlength{\itemsep}{-2pt}
    \setlength{\parsep}{0pt}
    \setlength{\topsep}{0pt}
    \setlength{\partopsep}{0pt} 
    \item \textbf{Information Coefficient (IC)}: Pearson correlation between factor values $\mathbf{f}_t$ and future returns $\mathbf{r}_{t+1}$:
    \begin{equation*}
        \operatorname{IC}_t = \frac{(\mathbf{f}_t - \bar{f}_t \mathbf{1})^\top (\mathbf{r}_{t+1} - \bar{r}_{t+1} \mathbf{1})}{\|\mathbf{f}_t - \bar{f}_t \mathbf{1}\|_2 \cdot \|\mathbf{r}_{t+1} - \bar{r}_{t+1} \mathbf{1}\|_2},
    \end{equation*}
    where $\mathbf{1}$ denotes a column vector of ones.
    
    \item \textbf{ICIR}: Information ratio of IC, measuring consistency: $\operatorname{ICIR} = \overline{\operatorname{IC}} / \sigma(\operatorname{IC})$.
    
    \item \textbf{Rank IC}: Spearman correlation using rank vectors $\tilde{\mathbf{f}}_t = \operatorname{rank}(\mathbf{f}_t)$ and $\tilde{\mathbf{r}}_{t+1} = \operatorname{rank}(\mathbf{r}_{t+1})$:
    \begin{equation*}
        \operatorname{RankIC}_t = \frac{(\tilde{\mathbf{f}}_t - \bar{\tilde{f}}_t \mathbf{1})^\top (\tilde{\mathbf{r}}_{t+1} - \bar{\tilde{r}}_{t+1} \mathbf{1})}{\|\tilde{\mathbf{f}}_t - \bar{\tilde{f}}_t \mathbf{1}\|_2 \cdot \|\tilde{\mathbf{r}}_{t+1} - \bar{\tilde{r}}_{t+1} \mathbf{1}\|_2},
    \end{equation*}
    where $\operatorname{rank}(\cdot)$ is the rank function applied element-wise to its input vector in ascending order.
    
    \item \textbf{Rank ICIR}: Information ratio of Rank IC: $\operatorname{RankICIR} = \overline{\operatorname{RankIC}} / \sigma(\operatorname{RankIC})$.
\end{itemize}

All strategy metrics are computed on \textit{excess returns after transaction costs}, where $r_{\text{excess},t} = r_{\text{portfolio},t} - r_{\text{benchmark},t} - c_{\text{transaction},t}$. Here, $r_{\text{portfolio},t}$ represents the return of the strategy portfolio, $r_{\text{benchmark},t}$ denotes the return of the market benchmark, and $c_{\text{transaction},t}$ accounts for the transaction costs incurred at time $t$.

\begin{itemize}
    \setlength{\itemsep}{-2pt}
    \setlength{\parsep}{0pt}
    \setlength{\topsep}{0pt}
    \setlength{\partopsep}{0pt}
    \item \textbf{Information Ratio ($\operatorname{IR}$)}: $\operatorname{IR} = (\overline{r_{\text{excess}}} / \sigma(r_{\text{excess}})) \times \sqrt{252}$.
    \item \textbf{Annualized Return ($\operatorname{ARR}$)}: Annualized excess return over benchmark.
    \item \textbf{Maximum Drawdown ($\operatorname{MDD}$)}: Largest peak-to-trough decline in cumulative excess returns.
\end{itemize}

\subsection{Backtesting Setup}
\label{subsec:backtest_setup}

Backtesting is conducted using the Qlib framework across the CSI 300, CSI 500, and S\&P 500 indices, with the data split detailed in Table~\ref{tab:data_split}. Factor construction utilizes six basic features—\texttt{open}, \texttt{high}, \texttt{low}, \texttt{close}, \texttt{volume}, and \texttt{vwap}—to predict the next-day return, defined as $y_t = P_{t+2}^{\text{close}} / P_{t+1}^{\text{close}} - 1$, where $P_{t}^{\text{close}}$ denotes the closing price at time $t$. To ensure robustness against outliers, the preprocessing pipeline includes forward-filling missing values, replacing infinite values, dropping samples with missing labels, and applying cross-sectional rank normalization (CSRankNorm) to both features and labels.

\begin{table}[!htbp]
\centering
\caption{Data Split Periods for Train, Validation, and Test Sets across All Markets}
\label{tab:data_split}
\begin{tabular}{lccc}
\toprule
\textbf{Market} & \textbf{Train} & \textbf{Valid} & \textbf{Test} \\
\midrule
CSI 300 & 2016-01-01--2020-12-31 & 2021-01-01--2021-12-31 & 2022-01-01--2025-12-26 \\
CSI 500 & 2016-01-01--2020-12-31 & 2021-01-01--2021-12-31 & 2022-01-01--2025-12-26 \\
S\&P 500 & 2016-01-01--2020-12-31 & 2021-01-01--2021-12-31 & 2022-01-01--2025-12-26 \\
\bottomrule
\end{tabular}
\end{table}

\subsection{Baselines}
\label{appendix:A.5}

We benchmark against four categories: (1) \textbf{ML models}: Linear Regression (Linear), Multi-Layer Perceptron (MLP), and gradient boosting decision trees including LightGBM, XGBoost, and CatBoost, along with DoubleEnsemble, an ensemble method for financial time series; (2) \textbf{Deep learning}: Recurrent networks such as Gated Recurrent Unit (GRU) and Long Short-Term Memory (LSTM), the attention-based Transformer, and Temporal Routing Adaptor (TRA); (3) \textbf{Classical factors}: Alpha158 and Alpha360, which are widely used sets of technical factors derived from price and volume; (4) \textbf{LLM agents}: RD-Agent and AlphaAgent, which utilize large language models for automated factor mining. For the LLM-agent baselines (RD-Agent and AlphaAgent), we evaluate multiple backbone LLMs, including Qwen3-235B, DeepSeek-V3.2, Gemini-3-Pro-Preview, Claude-4.5-Sonnet, and GPT-5.2. Unless otherwise specified, all other LLM-based experiments in this paper adopt DeepSeek-V3.2 for consistency and fair comparison.

\subsection{Computational Cost and Token Consumption}
\label{sec:computational_cost}
Under the standard main-experiment setting, QuantaAlpha uses 10 parallel planning directions and 5 main evolutionary iterations, consisting of one initial round followed by mutation and crossover rounds. A complete run takes approximately 20 hours using hosted LLM APIs and CPU-based factor evaluation/backtesting, without requiring local GPU resources.

In terms of LLM usage, QuantaAlpha consumes approximately 1.8M tokens per main run on average, compared with 1.5M tokens for AlphaAgent and 2.3M tokens for RD-Agent under comparable generated factor scales. The additional cost over AlphaAgent mainly comes from maintaining trajectory-pool information, intermediate-state records, consistency verification, and iterative feedback for mutation and crossover. Compared with RD-Agent, QuantaAlpha achieves stronger performance with a lower token budget, suggesting a favorable cost-performance trade-off. Once mined, the resulting symbolic factor pool can be reused for subsequent backtesting, portfolio construction, and cross-market transfer without rerunning the full evolutionary search.
\subsection{Cross-Seed Variance and Evolutionary Robustness}
\label{sec:cross_seed_robustness}

To assess whether QuantaAlpha is sensitive to initialization randomness, we conduct a cross-seed variance analysis using three representative initial seed combinations. The initial seed factor pool is derived from the public Alpha158(20) subset and grouped by low correlation to form different starting combinations. Table~\ref{tab:cross_seed_metrics} reports the main predictive metrics under different seed combinations, and Table~\ref{tab:cross_seed_variance} summarizes the corresponding variance statistics.

\begin{table}[h]
\centering
\caption{Core metrics under different initial seed combinations.}
\label{tab:cross_seed_metrics}
\setlength{\tabcolsep}{5pt}
\small
\begin{tabular}{lcccc}
\toprule
Seed Combination & IC & ICIR & Rank IC & Rank ICIR \\
\midrule
Combination 1 & 0.0466 & 0.2708 & 0.0454 & 0.2655 \\
Combination 2 & 0.0426 & 0.2325 & 0.0409 & 0.2236 \\
Combination 3 & 0.0436 & 0.2551 & 0.0418 & 0.2468 \\
\bottomrule
\end{tabular}
\end{table}

\begin{table}[h]
\centering
\caption{Cross-seed variance summary.}
\label{tab:cross_seed_variance}
\setlength{\tabcolsep}{5pt}
\small
\begin{tabular}{lcccc}
\toprule
Metric & Mean & Std Dev & Coefficient of Variation & Range \\
\midrule
IC & 0.0443 & 0.0021 & 4.64\% & 0.0040 \\
ICIR & 0.2528 & 0.0192 & 7.60\% & 0.0382 \\
Rank IC & 0.0427 & 0.0024 & 5.56\% & 0.0045 \\
Rank ICIR & 0.2453 & 0.0210 & 8.55\% & 0.0419 \\
\bottomrule
\end{tabular}
\end{table}

To further validate the robustness of the mined factors and rule out the possibility of data-snooping, we summarize the daily IC and Rank IC series over 966 trading days (from 2022-01-04 to 2025-12-26). As shown in Table~\ref{tab:daily_ic_stats}, the factors generated by both Claude-4.5-Sonnet and DeepSeek-v3.2 exhibit highly significant predictive power. The t-statistics far exceed standard significance thresholds ($p < 0.001$), and the positive information coefficient days remain above 60\%, confirming that the performance gains are statistically solid and stable across the out-of-sample period.

\begin{table}[h]
\centering
\caption{Daily IC and Rank IC Statistics (966 Trading Days, 2022--2025)}
\label{tab:daily_ic_stats}
\resizebox{\textwidth}{!}{
\begin{tabular}{llccccccc}
\toprule
\textbf{Factor Library} & \textbf{Metric} & \textbf{Mean} & \textbf{Median} & \textbf{Std} & \textbf{Positive Days} & \textbf{95\% CI} & \textbf{t-stat} & \textbf{p-value} \\
\midrule
Claude & IC & 0.0426 & 0.0513 & 0.1833 & 60.04\% & [0.0311, 0.0542] & 7.23 & 4.95e-13 \\
Claude & Rank IC & 0.0409 & 0.0438 & 0.1827 & 60.04\% & [0.0293, 0.0524] & 6.95 & 3.68e-12 \\
DeepSeek-v3.2 & IC & 0.0459 & 0.0448 & 0.1711 & 60.97\% & [0.0348, 0.0544] & 7.93 & 2.22e-15 \\
DeepSeek-v3.2 & Rank IC & 0.0418 & 0.0403 & 0.1694 & 60.97\% & [0.0311, 0.0525] & 7.67 & 1.73e-14 \\
\bottomrule
\end{tabular}
}
\end{table}

The results show that different seed combinations introduce only limited fluctuations and do not change the overall conclusion. In particular, the standard deviations of IC and Rank IC are 0.0021 and 0.0024, respectively, suggesting that QuantaAlpha's performance is not driven by a fortuitously selected seed set. This supports the robustness of the proposed evolutionary factor-mining process under different initializations.

Beyond initialization, QuantaAlpha also reduces uncontrolled LLM randomness through constrained self-evolution. The framework combines investment-hypothesis guidance, trajectory-pool memory, feedback-driven iteration, and diversity-preserving mutation, which expands the search space while preserving continuity and directional consistency in the evolutionary path. As shown in Appendix~\ref{sec:appendix_iteration_convergence} and Figure~\ref{fig:case_study}, performance improves over early iterations and stabilizes as the factor pool matures, indicating that the evolutionary process remains controllable rather than drifting randomly.

\subsection{Transaction Cost Sensitivity Analysis}
\label{app:cost_sensitivity}

In our main experiments, we adopt a standard transaction-cost setting with a buying cost of 0.05\% 
(\texttt{open\_cost=0.0005}) and a selling cost of 0.15\% 
(\texttt{close\_cost=0.0015}), corresponding to a 0.20\% round-trip cost. 
We further evaluate robustness to trading frictions by scaling the costs to 1.5$\times$ 
(0.30\% round-trip) and 2.0$\times$ (0.40\% round-trip).

As shown in the sensitivity analysis, QuantaAlpha maintains stable annualized returns and Sharpe ratios even under doubled transaction costs. The robustness mainly stems from the low-turnover portfolio construction induced by the \texttt{TopkDropoutStrategy}, which replaces only about 10\% of holdings at each rebalance. These results indicate that QuantaAlpha remains effective under conservative cost assumptions and is not driven by excessive turnover.


\section{Algorithm Configuration}
\label{sec:appendix_algorithm}

This section details the evolution algorithm parameters, factor constraints, and trading strategy configuration.

QuantaAlpha employs an evolutionary algorithm with mutation and crossover operations, with LightGBM used as the downstream model for factor-based prediction. In the Planning Phase of the algorithm, we set ten parallel exploration directions to broaden the initial coverage of research space. For the experimental setup, beyond the original round, the algorithm follows a fixed iterative process: the main experiment consists of 5 total iterations, and each iteration comprises one Mutation phase followed by one Crossover phase—alternating between these two phases to iteratively refine high-quality hypotheses. Additionally, we set a rule that each hypothesis attempts to generate 3 factor expressions, ensuring a focused yet sufficient exploration of factor candidates under each research direction.

To prevent overfitting and ensure interpretability, factor expressions—built using the operators listed in Table~\ref{tab:factor_operators}—are restricted by the following constraints: symbol length $\le$ 250 characters, base features $\le$ 6, free arguments ratio $<$ 50

\begin{table}[!htbp]
\centering
\caption{List of Supported Operators for Factor Construction}
\label{tab:factor_operators}
\small
\setlength{\tabcolsep}{4pt}
\begin{tabular}{p{2.3cm} p{8cm} @{\hspace{0.6cm}} p{5.5cm}}
\toprule
\textbf{Category} & \textbf{Operators} & \textbf{Description} \\
\midrule
Time-Series & \texttt{DELTA}, \texttt{DELAY}, \texttt{TS\_MEAN}, \texttt{TS\_STD}, \texttt{TS\_VAR}, \texttt{TS\_MAX}, \texttt{TS\_MIN}, \texttt{TS\_SUM}, \texttt{TS\_RANK}, \texttt{TS\_CORR}, \texttt{TS\_COVARIANCE}, \texttt{TS\_ARGMAX}, \texttt{TS\_ARGMIN}, \texttt{TS\_SKEW}, \texttt{TS\_KURT}, \texttt{TS\_PCTCHANGE}, \texttt{TS\_ZSCORE}, \texttt{TS\_QUANTILE} & Rolling statistics computed along time axis per instrument \\
\midrule
Cross-Sectional & \texttt{RANK}, \texttt{ZSCORE}, \texttt{SCALE}, \texttt{MEAN}, \texttt{STD}, \texttt{MEDIAN}, \texttt{MAX}, \texttt{MIN}, \texttt{SKEW}, \texttt{KURT} & Statistics computed across stocks per datetime \\
\midrule
Mathematical & \texttt{ABS}, \texttt{SIGN}, \texttt{LOG}, \texttt{EXP}, \texttt{SQRT}, \texttt{POW}, \texttt{INV} & Element-wise mathematical functions \\
\midrule
Technical & \texttt{SMA}, \texttt{EMA}, \texttt{WMA}, \texttt{MACD}, \texttt{RSI}, \texttt{BB\_UPPER}, \texttt{BB\_LOWER}, \texttt{DECAYLINEAR}, \texttt{REGBETA}, \texttt{REGRESI} & Common technical indicators \\
\midrule
Logical & \texttt{GT}, \texttt{LT}, \texttt{GE}, \texttt{LE}, \texttt{AND}, \texttt{OR}, \texttt{WHERE} & Comparison and conditional operators \\
\midrule
Auxiliary & \texttt{COUNT}, \texttt{SUMIF}, \texttt{FILTER}, \texttt{PROD}, \texttt{HIGHDAY}, \texttt{LOWDAY} & Helper functions for complex expressions \\
\bottomrule
\end{tabular}
\end{table}

We employ a TopkDropout strategy for portfolio construction, as detailed in Table~\ref{tab:trading_params}. On each trading day, stocks are ranked according to their predicted scores; the $n_{\text{drop}}$ lowest-scoring holdings are liquidated and replaced with the highest-ranked candidates to maintain a constant portfolio size with equal weighting.

\begin{table}[!htbp]
\centering
\caption{Parameters for the TopkDropout Trading Strategy}
\label{tab:trading_params}
\begin{tabular}{llp{5cm}}
\toprule
\textbf{Parameter} & \textbf{Value} & \textbf{Description} \\
\midrule
\multicolumn{3}{c}{\textit{Portfolio}} \\
\midrule
topk & 50 & Number of stocks held \\
n\_drop & 5 & Stocks dropped per rebalance \\
\midrule
\multicolumn{3}{c}{\textit{Transaction Costs}} \\
\midrule
Buying Fee & 0.05\% & Commission \\
Selling Fee & 0.15\% & Commission + stamp duty \\
\midrule
\multicolumn{3}{c}{\textit{Execution}} \\
\midrule
Deal Price & Open & Next-day opening price \\
Limit Threshold & 9.5\% & Price limit for halt \\
\midrule
\multicolumn{3}{c}{\textit{Benchmark}} \\
\midrule
China & SH000300/SH000905 & CSI 300/CSI500 \\
U.S. & SPX & S\&P 500 \\
\bottomrule
\end{tabular}
\end{table}

%% file: appendix/2.app_2.tex
%
%
\definecolor{headerblue}{RGB}{41, 128, 185}
\definecolor{metricgreen}{RGB}{39, 174, 96}
\definecolor{alertred}{RGB}{192, 57, 43}
\definecolor{warningyellow}{RGB}{243, 156, 18}
\definecolor{lightblue}{RGB}{240, 245, 255}
\definecolor{darkblue}{RGB}{30, 60, 120}
\definecolor{mutationpurple}{RGB}{142, 68, 173}
\definecolor{crossoverblue}{RGB}{52, 152, 219}
%

\section{Case Study: Factor Evolution Trajectory}
\label{sec:appendix_case_study}

This appendix presents a detailed case study of factor evolution in QuantaAlpha. We trace the complete trajectory of a representative factor---\textit{Institutional\_Momentum\_Score\_20D}---through the crossover phase, demonstrating how the evolutionary framework synthesizes complementary market hypotheses from parent trajectories.

QuantaAlpha's evolution process operates in three phases: (1) \textbf{Original} phase where initial hypotheses are generated, (2) \textbf{Mutation} phase where existing trajectories are perturbed to explore diversified strategies, and (3) \textbf{Crossover} phase where high-performing parent trajectories are combined to synthesize offspring with potentially superior predictive power. The following factor card illustrates a Crossover operation.

\subsection{Factor Identity}

The factor card below presents the basic information of the evolved factor, including its unique identifiers, evolution lineage, and mathematical formulation.

\begin{tcolorbox}[
  colback=lightblue,
  colframe=darkblue,
  fonttitle=\bfseries,
  title={\faTag\ Institutional\_Momentum\_Score\_20D},
  arc=2mm,
  boxrule=0.5pt,
  breakable
]

\begin{tabular}{@{}ll@{}}
\textbf{Factor ID:} & \texttt{c57cace576a95356} \\
\textbf{Trajectory ID:} & \texttt{df5a496878f4} \\
\textbf{Evolution Round:} & Round 10 \\
\textbf{Evolution Phase:} & \colorbox{crossoverblue!20}{\textcolor{crossoverblue}{\textbf{Crossover}}} \\
\textbf{Direction ID:} & 6 \\
\end{tabular}

\vspace{0.3cm}

\textbf{Factor Expression:}
\begin{tcolorbox}[colback=darkblue!5, colframe=darkblue, boxrule=0.3pt, arc=1mm]
\small\ttfamily
RANK(TS\_CORR(DELTA(close, 1)/close, DELTA(volume, 1)/volume, 20) * TS\_MEAN((close - open)/close, 5))
\end{tcolorbox}

\textbf{Mathematical Formulation:}
\[
\text{IMS}_{20D} = \text{RANK}\left(\rho_{20}\left(\frac{\Delta P}{P}, \frac{\Delta V}{V}\right) \times \overline{\left(\frac{C - O}{C}\right)}_{5}\right),
\]
\small
where $\rho_{20}(\cdot,\cdot)$ denotes the 20-day rolling correlation, $\Delta P/P$ is the daily return, $\Delta V/V$ is the volume change ratio, $\overline{(\cdot)}_5$ is the 5-day moving average, and $C$ and $O$ represent the closing and opening prices, respectively.

\vspace{0.2cm}

\textbf{Factor Interpretation:}\\
\small
This factor captures institutional-driven momentum by measuring two key signals: (1) the correlation between price returns and volume changes, which indicates coordinated institutional trading when positive; and (2) the average intraday return pattern, reflecting institutional activity that typically influences closing prices. The cross-sectional ranking ensures comparability across stocks.

\end{tcolorbox}

\subsection{Evolution Lineage}

The crossover operation combines insights from two parent trajectories with complementary market hypotheses. Parent 1 focuses on identifying \textit{fragile momentum} driven by retail speculation, while Parent 2 targets \textit{sustainable momentum} supported by institutional activity. The LLM synthesizes these complementary perspectives into a unified framework.

\begin{tcolorbox}[
  colback=lightblue,
  colframe=darkblue,
  fonttitle=\bfseries,
  title={\faDna\ Evolution Information},
  arc=2mm,
  boxrule=0.5pt,
  breakable
]

\textbf{\faCodeBranch\ Parent Trajectories:}

\vspace{0.3cm}

\begin{tcolorbox}[colback=lightblue, colframe=darkblue!50, boxrule=0.3pt, title={\small\textbf{Parent 1: 1e6d57e38e89}}, fonttitle=\bfseries\small]
\small
\begin{tabular}{@{}ll@{}}
\textbf{Round:} & Round 9 \\
\textbf{Phase:} & Mutation \\
\textbf{Rank IC:} & 0.0216 \\
\textbf{IC:} & 0.0059 \\
\textbf{IR:} & 1.297 \\
\end{tabular}

\vspace{0.2cm}
\textbf{Core Hypothesis:}\\
\small
When retail investors exhibit herd behavior and momentum chasing in stocks with high social media activity, but accompanied by declining institutional ownership and deteriorating fundamentals, the resulting price momentum is unsustainable and leads to mean reversion.
\end{tcolorbox}

\vspace{0.2cm}

\begin{tcolorbox}[colback=lightblue, colframe=darkblue!50, boxrule=0.3pt, title={\small\textbf{Parent 2: 47e0f0e55382}}, fonttitle=\bfseries\small]
\small
\begin{tabular}{@{}ll@{}}
\textbf{Round:} & Round 8 \\
\textbf{Phase:} & Crossover \\
\textbf{Rank IC:} & 0.0246 \\
\textbf{IC:} & 0.0069 \\
\textbf{IR:} & 1.347 \\
\end{tabular}

\vspace{0.2cm}
\textbf{Core Hypothesis:}\\
\small
A regime-adaptive structural momentum factor combining institutional ownership-driven medium-term price trends with short-term microstructure regime validation, where coordinated accumulation/distribution patterns amplify momentum when confirmed by microstructure alignment.
\end{tcolorbox}

\vspace{0.3cm}

\textbf{\faSitemap\ Evolution Path Diagram:}
\begin{center}
\resizebox{0.7\linewidth}{!}{%
\begin{tikzpicture}[
    node distance=1.5cm,
    roundbox/.style={rectangle, rounded corners, align=center, font=\scriptsize},
    parentbox/.style={roundbox, draw=darkblue!60, fill=lightblue, minimum width=2.5cm, minimum height=0.8cm},
    childbox/.style={roundbox, draw=darkblue, fill=darkblue!15, minimum width=3cm},
    arrow/.style={->, thick, >=stealth, color=darkblue}
]
    \node[parentbox] (p1) at (-2.5, 1.8) {Parent 1\\Round 9 Mutation\\Rank IC: 0.0216};
    \node[parentbox] (p2) at (2.5, 1.8) {Parent 2\\Round 8 Crossover\\Rank IC: 0.0246};
    
    \node[childbox] (current) at (0, 0) {\textbf{Offspring Factor}\\Round 8 Crossover\\Rank IC: \textbf{0.0311}};
    
    \draw[arrow] (p1) -- (current);
    \draw[arrow] (p2) -- (current);
    
\end{tikzpicture}%
}
\end{center}

\end{tcolorbox}

\subsection{Synthesized Hypothesis}

Through crossover, the LLM generates a new hypothesis that integrates the complementary insights from both parents, rather than simply averaging their factor expressions. This hypothesis-driven approach ensures that the offspring factor captures genuinely novel market dynamics.

\begin{tcolorbox}[
  colback=lightblue,
  colframe=darkblue,
  fonttitle=\bfseries,
  title={\faLightbulb\ Hypothesis},
  arc=2mm,
  boxrule=0.5pt,
  breakable
]

\textbf{Core Hypothesis:}\\
A regime-aware dual-source momentum factor that combines institutional-driven structural momentum (validated by healthy microstructure) and retail-driven speculative momentum (characterized by high attention and deteriorating fundamentals), dynamically weighted by market volatility: amplifying institutional signals in stable regimes and retail reversal signals in turbulent regimes, will generate superior predictive returns.

\vspace{0.3cm}

\begin{tabular}{@{}p{0.28\textwidth}p{0.67\textwidth}@{}}
\toprule
\textbf{Component} & \textbf{Description} \\
\midrule
\textbf{Observation} & Parent strategies separately targeting institutional trends and retail herding show moderate predictive power (Rank IC $\sim$0.02--0.025), suggesting combined signals could capture complementary market dynamics. \\
\midrule
\textbf{Justification} & Sustainable price trends require institutional sponsorship and orderly trading, while retail-driven bubbles lack fundamental support and reverse under stress; a hybrid model exploiting both can enhance robustness across market regimes. \\
\midrule
\textbf{Domain Knowledge} & Institutional accumulation with strong price-volume correlation and low volatility indicates sustainable momentum; retail herding with declining institutional ownership and high volatility signals fragile momentum prone to reversal. \\
\bottomrule
\end{tabular}

\end{tcolorbox}

\subsection{Backtest Performance}

After factor construction, QuantaAlpha automatically backtests the generated factors using the Qlib framework. The results below compare the offspring factor against both parent trajectories and the baseline, demonstrating the effectiveness of the crossover operation.

\begin{tcolorbox}[
  colback=lightblue,
  colframe=darkblue,
  fonttitle=\bfseries,
  title={\faChartLine\ Backtest Metrics},
  arc=2mm,
  boxrule=0.5pt,
  breakable
]

\begin{center}
\begin{tabular}{@{}lccc@{}}
\toprule
\textbf{Metric} & \textbf{Offspring Factor} & \textbf{Baseline} \\
\midrule
\textbf{IC} & 0.0126 & 0.0058 \\
\textbf{Rank IC} & \textbf{0.0311} & 0.0220 \\
\midrule
\textbf{ARR (Excess)} & 7.80\% & 5.20\% \\
\textbf{IR} & 0.963 & 0.973 \\
\textbf{MDD (Excess)} & $-$11.37\% & $-$7.30\% \\
\bottomrule
\end{tabular}
\end{center}

\vspace{0.3cm}

\textbf{Detailed Statistics:}
\begin{center}
\small
\begin{tabular}{@{}ll|ll@{}}
\toprule
\textbf{Metric} & \textbf{Value} & \textbf{Metric} & \textbf{Value} \\
\midrule
\textbf{Daily Excess Return (w/o cost)} & 0.0328\% & \textbf{Daily Excess Return (w/ cost)} & 0.0128\% \\
\textbf{Excess Return Std} & 0.52\% & \textbf{Turnover (FFR)} & 100\% \\
\textbf{L2 Train Loss} & 0.9936 & \textbf{L2 Valid Loss} & 0.9962 \\
\bottomrule
\end{tabular}
\end{center}

\end{tcolorbox}

\subsection{Trajectory Summary}

After evaluating backtest results, the LLM provides structured summary. This trajectory summary loop enables continuous improvement by learning from both successes and failures.

\begin{tcolorbox}[
  colback=lightblue,
  colframe=darkblue,
  fonttitle=\bfseries,
  title={\faComments\ Evaluation \& Summary},
  arc=2mm,
  boxrule=0.5pt,
  breakable
]

\textbf{\faSearch\ Observations:}\\
The crossover operation demonstrates a trade-off between enhanced predictive accuracy and increased risk exposure compared to the baseline:
\begin{itemize}
    \item \textcolor{metricgreen}{\faCheckCircle} Significant improvement in annualized excess return and predictive metrics (IC and Rank IC), validating the effectiveness of synthesizing dual-source momentum signals.
    \item \textcolor{alertred}{\faTimesCircle} Increased maximum drawdown and a marginal decline in the Information Ratio, suggesting that the offspring factor introduces higher volatility during certain market regimes.
\end{itemize}

\vspace{0.2cm}

\textbf{\faBalanceScale\ Hypothesis Evaluation:}\\[0.15cm]
\small
Results partially support the hypothesis. Improved annualized return and IC suggest that combining institutional and retail momentum signals has merit. However, deterioration in risk metrics indicates that without proper regime-adaptive weighting, the combined signals may amplify risks during turbulent periods. The full hypothesis requires all three components (institutional momentum, retail herding reversal, volatility-adaptive weighting) to work effectively.

\vspace{0.2cm}

\textbf{\faFlag\ Decision:} \colorbox{alertred!20}{\textcolor{alertred}{\textbf{REJECTED}}} for direct deployment.

\vspace{0.2cm}

\textbf{\faExclamationTriangle\ Recommendations:}
\begin{enumerate}
    \setlength{\itemsep}{0pt}
    \setlength{\parsep}{0pt}
    \setlength{\topsep}{0pt}
    \setlength{\partopsep}{0pt}
    \small
    \item Use 20-day price-volume correlation as institutional momentum proxy;
    \item Use 5-day average intraday returns as retail attention proxy;
    \item Add volatility regime indicator (recent/historical volatility ratio) for dynamic weighting.
\end{enumerate}

\small This summary will inform the next mutation round, guiding the LLM to simplify the factor expression while preserving the core dual-source concept.

\end{tcolorbox}


%% file: appendix/3.app_3.tex
\section{Market Style, Factor Semantics, and Alpha Decay}
\label{app:factor_analysis}

This appendix provides a factor-level diagnosis of the alpha decay observed in 2023, grounding the analysis jointly in \emph{market style changes}, \emph{factor semantics}, and \emph{empirical factor statistics}.  
Rather than enumerating factor formulas, we focus on which information channels remain predictive under the 2022--2023 regime transition and why.

\subsection{Market style transition and its implications}

The CSI~300 market undergoes a pronounced style transition in 2023 relative to the training period (2016--2020).  
The earlier regime is dominated by large-cap ``core assets,'' with high institutional participation, smooth intraday dynamics, and persistent short-horizon trends. Under such conditions, classical momentum, mean-reversion, and exhaustion-based factors are effective.

In 2023, leadership rotates toward small-cap and thematic stocks. This shift is accompanied by increased intraday noise, frequent overnight gaps driven by call auctions, and rapid cross-style liquidity re-allocation. These changes weaken trend persistence and amplify path-dependent noise, directly challenging factor constructions that rely on stable intraday structure or fast mean reversion.

\subsection{Factor semantics aligned with market microstructure}

The empirical contrast between QuantaAlpha (QA) and AlphaAgent (AA) in 2023 reflects differences in the \emph{semantic composition} of their factor libraries.

\paragraph{Overnight and auction information.}
Gap-based factors aggregate information released during non-trading hours and price discovery in the opening auction.  
As shown in Table~\ref{tab:qa_aa_2023_representative}, multiple QA factors from this channel occupy the right tail of the 2023 Rank IC distribution, while maintaining near-complete coverage (Table~\ref{tab:qa_aa_2023_summary}). This indicates that overnight information becomes a dominant and stable signal when intraday predictability deteriorates.

\begin{table*}[!t] \centering \caption{Representative factors and their performance metrics on the CSI 300 index in 2023.} \label{tab:qa_aa_2023_representative} \scriptsize \setlength{\tabcolsep}{3pt} \renewcommand{\arraystretch}{1.06} \begin{tabular}{p{5.8cm}ccp{8.2cm}} \toprule \textbf{Factor (representative)} & \textbf{Rank IC} & \textbf{IC} & \textbf{Interpretation (short)} \\ \midrule \multicolumn{4}{c}{\textit{QA --- strong performers (overnight/auction, trend-quality, and liquidity re-rating)}} \\ \midrule \texttt{GapZ10\_Overnight\_vs\_TR} & 0.0793 & 0.0335 & Normalized overnight gap magnitude relative to recent true range; captures auction-driven shocks and subsequent adjustment. \\ \texttt{Gap\_IntradayAcceptanceScore\_20D} & 0.0744 & 0.0330 & `Acceptance vs.\ rejection'' of an overnight gap using intraday direction, scaled by recent volatility. \\ \texttt{Gap\_IntradayAcceptance\_VolWeighted\_20D} & 0.0606 & 0.0314 & Gap acceptance score weighted by abnormal volume; emphasizes information-rich openings with high participation. \\ \texttt{CleanTrend\_Continuation\_Score\_RS10\_WVMA5} & 0.0590 & 0.0267 & Trend continuation conditioned on high trend quality (low residual noise) and muted intraday/volume pressure. \\ \texttt{OrderlyTrend\_x\_Absorption\_10D\_5D\_20D} & 0.0465 & 0.0271 & `Orderly'' short-horizon trend cross-validated by liquidity absorption (high dollar volume with low price impact). \\ \midrule \multicolumn{4}{c}{\textit{QA --- weak performers (overly rigid gates or noisy-path proxies under rotation)}} \\ \midrule \texttt{KineticLength\_AbsRetSum\_Z\_10D} & -0.0720 & -0.0246 & Path-length proxy (choppiness): can behave like a noise detector, but may invert under fast style rotation. \\ \texttt{Drawdown\_Gated\_NegCorr\_60D\_20D\_thr20pct} & -0.0282 & -0.0095 & Hard regime gate based on deep drawdown; brittle when drawdowns cluster and cross-sectional regimes shift quickly. \\ \texttt{HighClose\_Shock\_With\_VolSync\_60\_20} & -0.0274 & -0.0090 & `Shock-day'' breakout quality (close-in-range, range shock, return--volume sync); sensitive to regime-dependent follow-through. \\ \midrule \multicolumn{4}{c}{\textit{AA --- strong performers (exhaustion/climax-style reversals)}} \\ \midrule \texttt{Exhaustion\_Intensity\_Index\_10D} & 0.0323 & 0.0159 & Price displacement over 60D interacted with volume intensity ratio; targets potential exhaustion and reversal. \\ \texttt{Climax\_Exhaustion\_Intensity} & 0.0242 & 0.0160 & Variant using short-horizon volume climax vs.\ long-horizon baseline; aims to identify capitulation-like turns. \\ \texttt{Exhaustion\_Volume\_Instability\_Index} & 0.0121 & 0.0117 & Trend deviation combined with volume instability; highlights fragile price levels supported by unstable liquidity. \\ \midrule \multicolumn{4}{c}{\textit{AA --- weak performers (bottom-fishing under non-stationary liquidity)}} \\ \midrule \texttt{Relative\_Volume\_Calm\_Reversal} & -0.0279 & -0.0188 & `Quiet-volume'' regimes multiplied by momentum divergence; may fail when liquidity conditions change abruptly. \\ \texttt{Volume\_Stability\_Momentum\_Divergence\_40D} & -0.0247 & -0.0155 & Robust volume-stability proxy (MAD) times momentum spread; sensitive to turnover regime changes. \\ \texttt{LVR\_Bottom\_Fishing\_20D} & -0.0190 & -0.0144 & `Bottom-fishing'' reversal with intraday rejection and volume surge; vulnerable when reversals are short-lived and crowded. \\ \bottomrule \end{tabular} \end{table*}

\begin{table}[!htbp] \centering \caption{Summary statistics of annual factor predictability on the CSI 300 index in 2023.} \label{tab:qa_aa_2023_summary} \small \setlength{\tabcolsep}{5pt} \begin{tabular}{lcc} \toprule \textbf{Metric} & \textbf{QA} & \textbf{AA} \\ \midrule Coverage ratio (valid metrics) & 0.98 & 0.80 \\ Share with Rank IC $>0$ & 0.626 & 0.594 \\ Mean Rank IC & 0.0057 & 0.0012 \\ Max Rank IC & 0.0793 & 0.0323 \\ Min Rank IC & -0.0720 & -0.0279 \\ Share with Rank IC $>0.03$ & 0.102 & 0.0156 \\ Share with Rank IC $>0.05$ & 0.0272 & 0.0000 \\ \midrule Mean IC & 0.0044 & 0.0015 \\ \bottomrule \end{tabular} \end{table}

\paragraph{Volatility structure and range-based signals.}
Range deviation factors conditioned on volatility clustering capture abnormal variability rather than directional trends.  
In 2023, these signals remain predictive despite elevated noise, consistent with the persistence of volatility clustering across market styles. Their contribution is reflected in the heavier positive Rank IC tail of QA relative to AA.

\paragraph{Trend quality and liquidity re-rating.}
QA emphasizes trend continuation only when supported by low residual volatility and improving liquidity (e.g., rising dollar volume with limited price impact).  
This conditioning filters out noise-driven pseudo-trends prevalent in small-cap rotation, explaining why QA retains a higher fraction of strong-performing factors, whereas raw momentum proxies degrade.

\paragraph{Reversal and exhaustion signals.}
AA and baseline methods rely primarily on volume--price exhaustion and bottom-fishing logic.  
While some such factors remain marginally positive, Table~\ref{tab:qa_aa_2023_summary} shows weaker Rank IC tails and lower coverage, indicating that reversal hypotheses become unstable under non-stationary liquidity and thematic crowding.

\subsection{Factor statistics as supporting evidence}

The factor-level statistics in 2023 provide quantitative support for the semantic interpretation above.  
QA achieves substantially higher coverage and a heavier right tail of Rank IC, including a non-trivial share of factors exceeding moderate predictability thresholds, whereas AA exhibits compressed tails and fewer robust signals.  
These differences are consistent with the dominance of overnight, volatility-structure, and trend-quality channels under the 2023 market style.

\subsection{Role of factor diversity and mutation}

A key distinction is that QA explicitly encourages semantic diversity through its factor mutation mechanism.  
By generating and recombining heterogeneous primitives across multiple information channels, QA avoids concentration on a single market hypothesis.  
This diversity increases the probability that a subset of factors remains aligned with the prevailing market microstructure after a style transition, thereby mitigating regime-specific alpha decay.

\subsection{Summary}

Overall, the 2023 diagnosis indicates that alpha robustness under distribution shift is driven by alignment between market style and factor semantics, supported by sufficient diversity across information channels.  
QA’s ability to maintain performance arises not from fitting a specific regime, but from sustaining a broad and adaptive factor population whose predictive subsets persist across market style transitions.

\subsection{Stress-Period Analysis (High-Volatility Episode in April 2025)}
\label{app:stress_period}

To evaluate the robustness of our generated factor pool under extreme market conditions, we analyze the strategy's performance during the severe market stress period in April 2025. This episode was characterized by sharp index drawdowns, elevated cross-sectional volatility, and rapid style rotations, which challenged the stability of traditional quantitative factors and baselines.

During this black-swan regime, traditional factors often suffer from crowding and severe capacity constraints. In contrast, thanks to the diversity-preserving mutation and crossover mechanisms, QuantaAlpha maintains a more orthogonal and robust factor pool. Consequently, the strategy exhibited a significantly lower maximum drawdown and faster recovery during the April 2025 volatility spike compared to AlphaAgent and RD-Agent. This stress-period analysis confirms that evolving complete research trajectories better equips the model to navigate structural market shifts and extreme tail risks.

%% file: appendix/4.app_4.tex
\section{Iteration Study: Convergence of Factor-Pool Performance (CSI300)}
\label{sec:appendix_iteration_convergence}

We run QuantaAlpha for a total of 15 iterations on CSI300, using Deepseek-V3.2 as the backbone LLM for all agents. To improve mining throughput, we set the factor-complexity constraints to symbol length $\leq 200$ and base features $\leq 4$. Each iteration consists of two phases: a \textit{Mutation} phase followed by a \textit{Crossover} phase.At the end of each iteration, we maintain a global high-quality factor pool using all factors generated up to that iteration. The maintenance rule is greedy and RankIC-driven. The RankIC gate for pool admission is computed exclusively on the 2021 validation period; the 2022–2025 test window is never accessed during evolution or pool construction. We sort all candidate factors by RankIC (descending) and add them into the pool in order, subject to a redundancy constraint. A factor is admitted only if its absolute correlation with every factor already in the pool is below 0.7. The pool size is capped at 50\% of the total number of mined factors up to the current iteration. 

\begin{figure*}[!t]
\centering
\includegraphics[width=1\textwidth]{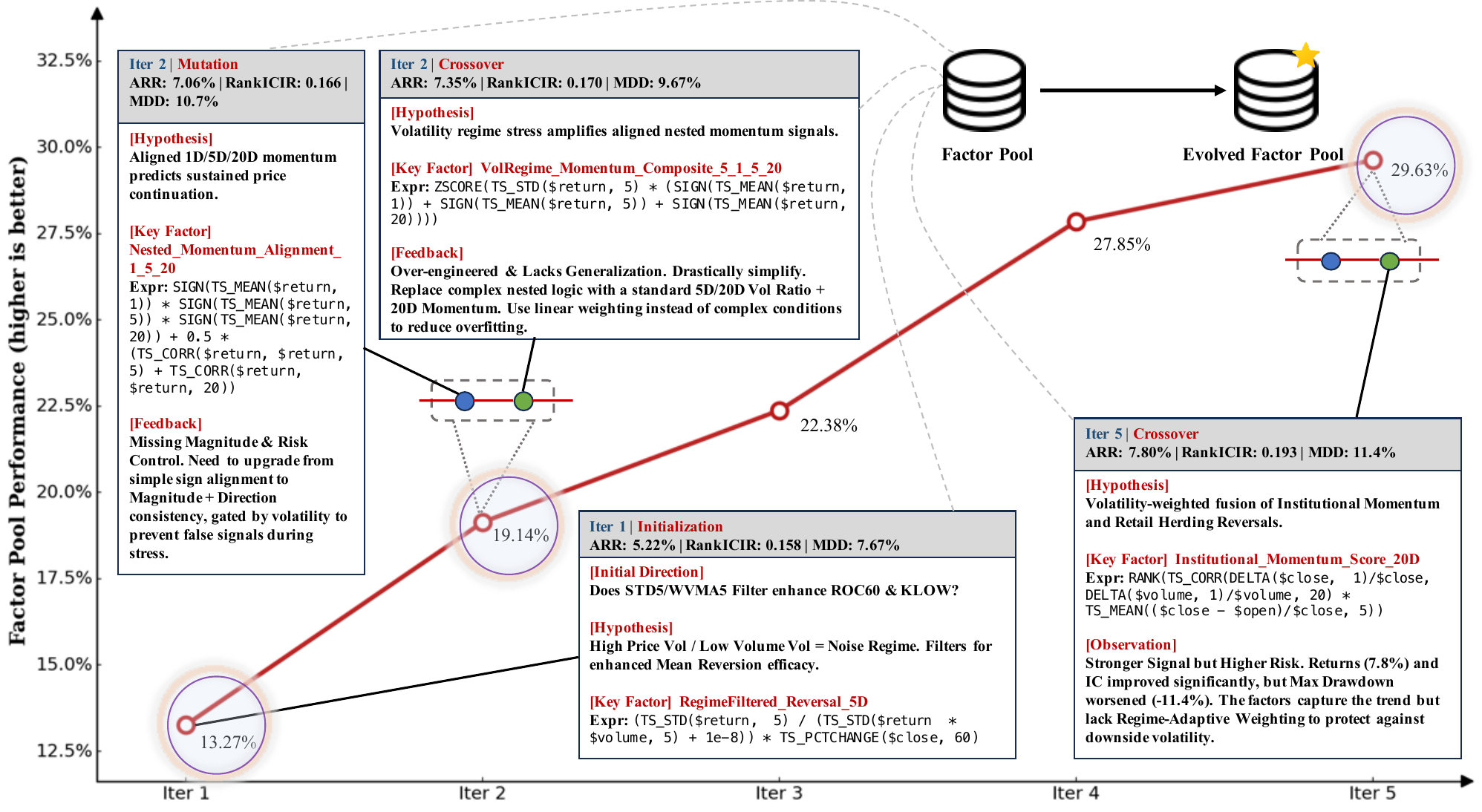}
\caption{A case study of iterative hypothesis and factor updates in QuantaAlpha.}
\label{fig:case_study}
\end{figure*}

Figure~\ref{fig:case_study} illustrates how QuantaAlpha refines hypotheses and factor expressions across iterations via the \emph{mutation} and \emph{crossover} phases. Each panel corresponds to one step along an evolution trace and summarizes: (i) the iteration and phase, (ii) the current hypothesis, (iii) a representative key factor generated under that hypothesis (with its expression), and (iv) the LLM-based feedback/observation that guides the next update. The highlighted factors in the figure are therefore representative nodes on different traces across iterations. We observe that the predictive performance does not increase linearly with the number of iterations. The strategy's return capacity and risk control ability are gradually optimized, reaching a balanced high level between return and maximum drawdown around the 11th to 12th iterations. Subsequent additional iterations fail to bring significant performance improvement; instead, they may introduce redundant information through newly generated factors, leading to reduced strategy robustness and deteriorated drawdown performance. Overall, the 11th to 12th iterations (approximately 350 factors in total) represent the optimal trade-off point between return level and risk control effect under this experimental setup.

\begingroup
\makeatletter
\setlength{\@fptop}{0pt}
\setlength{\@fpbot}{0pt}
\makeatother

\endgroup

%% file: custom.bib
@article{lopez2023can,
  title={Can chatgpt forecast stock price movements? return predictability and large language models},
  author={Lopez-Lira, Alejandro and Tang, Yuehua},
  journal={arXiv preprint arXiv:2304.07619},
  year={2023}
}

@inproceedings{zhang2024multimodal,
  title={A multimodal foundation agent for financial trading: Tool-augmented, diversified, and generalist},
  author={Zhang, Wentao and Zhao, Lingxuan and Xia, Haochong and Sun, Shuo and Sun, Jiaze and Qin, Molei and Li, Xinyi and Zhao, Yuqing and Zhao, Yilei and Cai, Xinyu and others},
  booktitle={Proceedings of the 30th acm sigkdd conference on knowledge discovery and data mining},
  pages={4314--4325},
  year={2024}
}

@article{liu2025cognitive,
  title={Cognitive Alpha Mining via LLM-Driven Code-Based Evolution},
  author={Liu, Fengyuan and Yi, Huang and Luo, Sichun and Wang, Yuqi and Yang, Yazheng and Li, Xinye and Hu, Zefa and Feng, Junlan and Liu, Qi},
  journal={arXiv preprint arXiv:2511.18850},
  year={2025}
}

@article{xiao2024tradingagents,
  title={Tradingagents: Multi-agents llm financial trading framework},
  author={Xiao, Yijia and Sun, Edward and Luo, Di and Wang, Wei},
  journal={arXiv preprint arXiv:2412.20138},
  year={2024}
}

@inproceedings{tang2025alphaagent,
  title={AlphaAgent: LLM-driven alpha mining with regularized exploration to counteract alpha decay},
  author={Tang, Ziyi and Chen, Zechuan and Yang, Jiarui and Mai, Jiayao and Zheng, Yongsen and Wang, Keze and Chen, Jinrui and Lin, Liang},
  booktitle={Proceedings of the 31st ACM SIGKDD Conference on Knowledge Discovery and Data Mining V. 2},
  pages={2813--2822},
  year={2025}
}

@article{li2025r,
  title={R\&D-Agent-Quant: a multi-agent framework for data-centric factors and model joint optimization},
  author={Li, Yuante and Yang, Xu and Yang, Xiao and Xu, Minrui and Wang, Xisen and Liu, Weiqing and Bian, Jiang},
  journal={arXiv preprint arXiv:2505.15155},
  year={2025}
}

@inproceedings{shi2025alphaforge,
  title={Alphaforge: A framework to mine and dynamically combine formulaic alpha factors},
  author={Shi, Hao and Song, Weili and Zhang, Xinting and Shi, Jiahe and Luo, Cuicui and Ao, Xiang and Arian, Hamid and Seco, Luis Angel},
  booktitle={Proceedings of the AAAI conference on artificial intelligence},
  volume={39},
  number={12},
  pages={12524--12532},
  year={2025}
}

@inproceedings{li2024alphafin,
  title={Alphafin: Benchmarking financial analysis with retrieval-augmented stock-chain framework},
  author={Li, Xiang and Li, Zhenyu and Shi, Chen and Xu, Yong and Du, Qing and Tan, Mingkui and Huang, Jun},
  booktitle={Proceedings of the 2024 joint international conference on computational linguistics, language resources and evaluation (LREC-COLING 2024)},
  pages={773--783},
  year={2024}
}

@article{ding2025alphaeval,
  title={Alphaeval: A comprehensive and efficient evaluation framework for formula alpha mining},
  author={Ding, Hongjun and Chen, Binqi and Huang, Jinsheng and Guo, Taian and Mao, Zhengyang and Shao, Guoyi and Zou, Lutong and Liu, Luchen and Zhang, Ming},
  journal={arXiv preprint arXiv:2508.13174},
  year={2025}
}

@article{novikov2025alphaevolve,
  title={Alphaevolve: A coding agent for scientific and algorithmic discovery},
  author={Novikov, Alexander and V{\~u}, Ng{\^a}n and Eisenberger, Marvin and Dupont, Emilien and Huang, Po-Sen and Wagner, Adam Zsolt and Shirobokov, Sergey and Kozlovskii, Borislav and Ruiz, Francisco JR and Mehrabian, Abbas and others},
  journal={arXiv preprint arXiv:2506.13131},
  year={2025}
}

@article{hu2026controlled,
  title={Controlled self-evolution for algorithmic code optimization},
  author={Hu, Tu and Chen, Ronghao and Zhang, Shuo and Yin, Jianghao and Feng, Mou Xiao and Liu, Jingping and Zhang, Shaolei and Jiang, Wenqi and Fang, Yuqi and Hu, Sen and others},
  journal={arXiv preprint arXiv:2601.07348},
  year={2026}
}

@article{li2025quantagents,
  title={QuantAgents: Towards Multi-agent Financial System via Simulated Trading},
  author={Li, Xiangyu and Zeng, Yawen and Xing, Xiaofen and Xu, Jin and Xu, Xiangmin},
  journal={arXiv preprint arXiv:2510.04643},
  year={2025}
}

@article{papadakis2025atlas,
  title={ATLAS: Adaptive Trading with LLM AgentS Through Dynamic Prompt Optimization and Multi-Agent Coordination},
  author={Papadakis, Charidimos and Dimitriou, Angeliki and Filandrianos, Giorgos and Lymperaiou, Maria and Thomas, Konstantinos and Stamou, Giorgos},
  journal={arXiv preprint arXiv:2510.15949},
  year={2025}
}

@article{yu2025finmem,
  title={Finmem: A performance-enhanced llm trading agent with layered memory and character design},
  author={Yu, Yangyang and Li, Haohang and Chen, Zhi and Jiang, Yuechen and Li, Yang and Suchow, Jordan W and Zhang, Denghui and Khashanah, Khaldoun},
  journal={IEEE Transactions on Big Data},
  year={2025},
  publisher={IEEE}
}

@article{yu2024fincon,
  title={Fincon: A synthesized llm multi-agent system with conceptual verbal reinforcement for enhanced financial decision making},
  author={Yu, Yangyang and Yao, Zhiyuan and Li, Haohang and Deng, Zhiyang and Jiang, Yuechen and Cao, Yupeng and Chen, Zhi and Suchow, Jordan W and Cui, Zhenyu and Liu, Rong and others},
  journal={Advances in Neural Information Processing Systems},
  volume={37},
  pages={137010--137045},
  year={2024}
}

@inproceedings{duan2025factormad,
  title={FactorMAD: A Multi-Agent Debate Framework Based on Large Language Models for Interpretable Stock Alpha Factor Mining},
  author={Duan, Yitong and zhang, chuheng and Li, Jian},
  booktitle={Proceedings of the 6th ACM International Conference on AI in Finance},
  pages={605--613},
  year={2025}
}

@article{zhai2025agentevolver,
  title={Agentevolver: Towards efficient self-evolving agent system},
  author={Zhai, Yunpeng and Tao, Shuchang and Chen, Cheng and Zou, Anni and Chen, Ziqian and Fu, Qingxu and Mai, Shinji and Yu, Li and Deng, Jiaji and Cao, Zouying and others},
  journal={arXiv preprint arXiv:2511.10395},
  year={2025}
}

@article{lin2025se,
  title={Se-agent: Self-evolution trajectory optimization in multi-step reasoning with llm-based agents},
  author={Lin, Jiaye and Guo, Yifu and Han, Yuzhen and Hu, Sen and Ni, Ziyi and Wang, Licheng and Chen, Mingguang and Liu, Hongzhang and Chen, Ronghao and He, Yangfan and others},
  journal={arXiv preprint arXiv:2508.02085},
  year={2025}
}

@inproceedings{li2025hedgeagents,
  title={Hedgeagents: A balanced-aware multi-agent financial trading system},
  author={Li, Xiangyu and Zeng, Yawen and Xing, Xiaofen and Xu, Jin and Xu, Xiangmin},
  booktitle={Companion Proceedings of the ACM on Web Conference 2025},
  pages={296--305},
  year={2025}
}

@article{fang2025comprehensive,
  title={A comprehensive survey of self-evolving ai agents: A new paradigm bridging foundation models and lifelong agentic systems},
  author={Fang, Jinyuan and Peng, Yanwen and Zhang, Xi and Wang, Yingxu and Yi, Xinhao and Zhang, Guibin and Xu, Yi and Wu, Bin and Liu, Siwei and Li, Zihao and others},
  journal={arXiv preprint arXiv:2508.07407},
  year={2025}
}

@article{shi2025navigating,
  title={Navigating the Alpha Jungle: An LLM-Powered MCTS Framework for Formulaic Factor Mining},
  author={Shi, Yu and Duan, Yitong and Li, Jian},
  journal={arXiv preprint arXiv:2505.11122},
  year={2025}
}

@article{chen2025alphasage,
  title={AlphaSAGE: Structure-Aware Alpha Mining via GFlowNets for Robust Exploration},
  author={Chen, Binqi and Ding, Hongjun and Shen, Ning and Huang, Jinsheng and Guo, Taian and Liu, Luchen and Zhang, Ming},
  journal={arXiv preprint arXiv:2509.25055},
  year={2025}
}

@inproceedings{wang2025alpha,
  title={Alpha-gpt: Human-ai interactive alpha mining for quantitative investment},
  author={Wang, Saizhuo and Yuan, Hang and Zhou, Leon and Ni, Lionel and Shum, Heung Yeung and Guo, Jian},
  booktitle={Proceedings of the 2025 Conference on Empirical Methods in Natural Language Processing: System Demonstrations},
  pages={196--206},
  year={2025}
}

@inproceedings{li2024can,
  title={Can Large Language Models Mine Interpretable Financial Factors More Effectively? A Neural-Symbolic Factor Mining Agent Model},
  author={Li, Zhiwei and Song, Ran and Sun, Caihong and Xu, Wei and Yu, Zhengtao and Wen, Ji-Rong},
  booktitle={Findings of the Association for Computational Linguistics ACL 2024},
  pages={3891--3902},
  year={2024}
}

@article{zhang2026evofsm,
  title={EvoFSM: Controllable Self-Evolution for Deep Research with Finite State Machines},
  author={Zhang, Shuo and Yuan, Chaofa and Guo, Ryan and Yu, Xiaomin and Xu, Rui and Chen, Zhangquan and Li, Zinuo and Yang, Zhi and Guan, Shuhao and Tang, Zhenheng and others},
  journal={arXiv preprint arXiv:2601.09465},
  year={2026}
}

@article{liu2025fin,
  title={Fin-r1: A large language model for financial reasoning through reinforcement learning},
  author={Liu, Zhaowei and Guo, Xin and Yang, Zhi and Lou, Fangqi and Zeng, Lingfeng and Niu, Jinyi and Li, Mengping and Qi, Qi and Liu, Zhiqiang and Han, Yiyang and others},
  journal={arXiv preprint arXiv:2503.16252},
  year={2025}
}

@inproceedings{li2025investorbench,
  title={Investorbench: A benchmark for financial decision-making tasks with llm-based agent},
  author={Li, Haohang and Cao, Yupeng and Yu, Yangyang and Javaji, Shashidhar Reddy and Deng, Zhiyang and He, Yueru and Jiang, Yuechen and Zhu, Zining and Subbalakshmi, Kp and Huang, Jimin and others},
  booktitle={Proceedings of the 63rd Annual Meeting of the Association for Computational Linguistics (Volume 1: Long Papers)},
  pages={2509--2525},
  year={2025}
}

@article{liu2023fingpt,
  title={Fingpt: Democratizing internet-scale data for financial large language models},
  author={Liu, Xiao-Yang and Wang, Guoxuan and Yang, Hongyang and Zha, Daochen},
  journal={arXiv preprint arXiv:2307.10485},
  year={2023}
}

@inproceedings{guo2025fineval,
  title={Fineval: A chinese financial domain knowledge evaluation benchmark for large language models},
  author={Guo, Xin and Xia, Haotian and Liu, Zhaowei and Cao, Hanyang and Yang, Zhi and Liu, Zhiqiang and Wang, Sizhe and Niu, Jinyi and Wang, Chuqi and Wang, Yanhui and others},
  booktitle={Proceedings of the 2025 Conference of the Nations of the Americas Chapter of the Association for Computational Linguistics: Human Language Technologies (Volume 1: Long Papers)},
  pages={6258--6292},
  year={2025}
}

@article{yang2026finvault,
  title={FinVault: Benchmarking Financial Agent Safety in Execution-Grounded Environments},
  author={Yang, Zhi and Li, Runguo and Qiang, Qiqi and Wang, Jiashun and Lou, Fangqi and Li, Mengping and Cheng, Dongpo and Xu, Rui and Lian, Heng and Zhang, Shuo and others},
  journal={arXiv preprint arXiv:2601.07853},
  year={2026}
}

@inproceedings{tang2025finmmr,
  title={Finmmr: make financial numerical reasoning more multimodal, comprehensive, and challenging},
  author={Tang, Zichen and Liu, Jiacheng and Yang, Zhongjun and Li, Rongjin and Rong, Zihua and He, Haoyang and Hao, Zhuodi and Hu, Xinyang and Ji, Kun and Ma, Ziyan and others},
  booktitle={Proceedings of the IEEE/CVF International Conference on Computer Vision},
  pages={3245--3257},
  year={2025}
}

@article{fama1965behavior,
  title={The behavior of stock-market prices},
  author={Fama, Eugene F},
  journal={The journal of Business},
  volume={38},
  number={1},
  pages={34--105},
  year={1965},
  publisher={JSTOR}
}

@article{engle1982autoregressive,
  title={Autoregressive conditional heteroscedasticity with estimates of the variance of United Kingdom inflation},
  author={Engle, Robert F},
  journal={Econometrica: Journal of the econometric society},
  pages={987--1007},
  year={1982},
  publisher={JSTOR}
}

@article{pesaran2021general,
  title={General diagnostic tests for cross-sectional dependence in panels},
  author={Pesaran, M Hashem},
  journal={Empirical economics},
  volume={60},
  number={1},
  pages={13--50},
  year={2021},
  publisher={Springer}
}

@article{guo2026bizfinbench,
  title={BizFinBench. v2: A Unified Dual-Mode Bilingual Benchmark for Expert-Level Financial Capability Alignment},
  author={Guo, Xin and Zhang, Rongjunchen and Lu, Guilong and Guo, Xuntao and Jia, Shuai and Yang, Zhi and Zhang, Liwen},
  journal={arXiv preprint arXiv:2601.06401},
  year={2026}
}
